\documentclass[a4paper,fleqn,usenatbib]{mnras}

\usepackage{newtxtext,newtxmath}

\usepackage[T1]{fontenc}
\usepackage{ae,aecompl}

\usepackage{graphicx}	
\usepackage{amsmath}	
\usepackage{amssymb}	
\usepackage{float}
\usepackage{subcaption}
\usepackage{bold-extra}

\title[Black hole-galaxy relations in {\sc{BlueTides}}]{{\sc{BlueTides}} simulation: 
establishing black hole-galaxy relations at high-redshift}

\author[K.-W. Huang et al.]{Kuan-Wei Huang,$^{1}$\thanks{E-mail: kuanweih@andrew.cmu.edu}
Tiziana Di Matteo,$^{1}$
Aklant K. Bhowmick,$^{1}$
Yu Feng$^{2}$
\newauthor
and Chung-Pei Ma$^{3}$
\\
$^{1}$McWilliams Center for Cosmology, Dept. of Physics, Carnegie Mellon University, Pittsburgh, PA, 15213, USA\\
$^{2}$Berkeley Center for Cosmological Physics, University of California at Berkeley, Berkeley, CA, 94720, USA\\
$^{3}$Dept. of Astronomy, University of California at Berkeley, Berkeley, CA, 94720, USA
}

\date{Accepted XXX. Received YYY; in original form ZZZ}

\pubyear{2018}

\begin{document}
\label{firstpage}
\pagerange{\pageref{firstpage}--\pageref{lastpage}}
\maketitle

\begin{abstract}
  The scaling relations between the mass of supermassive black holes
  ($M_{\bullet}$) and host galaxy properties (stellar mass,
  $M_{\star}$, and velocity dispersion, $\sigma$), provide a link
  between the growth of black holes (BHs) and that of their hosts.
  Here we investigate if and how the BH-galaxy relations are
  established in the high-$z$ universe using \textsc{BlueTides}, a
  high-resolution large volume cosmological hydrodynamic simulation.
  We find the $M_{\bullet}-M_{\star}$ and $M_{\bullet}-\sigma$
  relations at $z=8$:
  $\log_{10}(M_{\bullet}) = 8.25 + 1.10 \
  \log_{10}(M_{\star}/10^{11}M_{\sun})$
  and
  $\log_{10}(M_{\bullet}) = 8.35 + 5.31 \
  \log_{10}(\sigma/200kms^{-1})$
  at $z=8$, both fully consistent with the local measurements.  The
  slope of the $M_{\bullet}-\sigma$ relation is slightly steeper for
  high star formation rate and $M_{\star}$ galaxies while it remains
  unchanged as a function of Eddington accretion rate onto the BH.
  The intrinsic scatter in $M_{\bullet}-\sigma$ relation in all cases
  ($\epsilon \sim 0.4$) is larger at these redshifts than inferred from
  observations and larger than in $M_{\bullet}-M_{\star}$ relation
  ($\epsilon \sim 0.14$).  We find the gas-to-stellar ratio
  $f=M_{\rm gas}/M_{\star}$ in the host (which can be very high at
  these redshifts) to have the most significant impact setting the
  intrinsic scatter of $M_{\bullet}-\sigma$.  The scatter is
  significantly reduced when galaxies with high gas fractions
  ($\epsilon = 0.28$ as $f<10$) are excluded (making the sample more
  comparable to low-$z$ galaxies); these systems have the largest star
  formation rates and black hole accretion rates, indicating that
  these fast-growing systems are still moving toward the relation at
  these high redshifts. Examining the evolution (from $z=10$ to 8) of
  high mass black holes in $M_{\bullet}-\sigma$ plane confirms this trend.
\end{abstract}



\section{Introduction}
\label{sec:Intro}
Over the last three decades, scaling relations between mass of supermassive black holes (SMBHs) and several stellar properties of their host galaxies such as bulge stellar mass and bulge velocity dispersion \citep{Magorrian1998, Haring2004, Gebhardt2000, Tremaine2002, Gultekin2009, Kormendy2013, McConnell2013, Reines2015} have been discovered and measured for galaxies with black holes (BHs) and active galactic nuclei (AGN) from $z=0$ and up to $z \sim 2$ (using different techniques).

Many theoretical models have been developed to understand the origin of these relations. 
Several cosmological simulations that follow the formation, growth of BHs and their host galaxies have successfully reproduced the scaling relations at low-$z$; these include recent simulations such as the {\it Illustris} simulation \citep{Vogelsberger2014, Sijacki2015}, the Magneticum Pathfinder SPH simulation \citep{Steinborn2015}, the Evolution and Assembly of GaLaxies and their Environment ({\it EAGLE}) suite of SPH simulation \citep{Schaye2015}, and the {\it MassiveBlackII (MBII)} simulation \citep{Khandai2015, DeGraf2015}. 
Thus, the scaling relations from observations and simulations agree with each other at low-$z$, linking the growth of SMBHs to the growth of their hosts via AGN feedback.

A popular way to interpret the scaling relations is by invoking AGN feedback. 
Many models (and simulations) show that the SMBHs regulate their own growth with their hosts by coupling a fraction of their released energy back to the surrounding gas \citep{DiMatteo2005}. 
The BHs grow only until sufficient energy is released to unbind the gas from the local galaxy potential \citep{Silk1998, King2003, Springel2005, Bower2006, Croton2006, Dimatteo2008, Ciotti2009, Fanidakis2011}. 
However, there are also models which have been proposed to explain the scaling relations without invoking the foregoing coupled feedback mechanism. 
For instance, it has also been shown that dry mergers can potentially drive BHs and their hosts towards a mean relation \citep{Peng2007, Hirschmann2010, Jahnke2011}. 
Regardless, studying the scaling relations in both observation and simulation is essential for understanding the coupled growth of galaxies and BHs across cosmic history.

An important related question is when the scaling relations are established, and if they still persist at higher redshifts when the first massive BHs form ($z>6$). 
To understand this, galaxies with AGN play a key role in observations \citep{Bennert2010, Merloni2010, Kormendy2013}. 
A strong direct constraint on the high-redshift evolution of SMBHs comes from the luminous quasars at $z\sim6$ in SDSS \citep{Fan2006, Jiang2009, Mortlock2011}.
Most recently, the earliest quasar is discovered at $z=7.5$ \citep{Banados2017} in ALLWISE, UKIDSS, and DECaLS. 
However, it is still not established as to whether these objects follow the local BH-galaxy relations and whether there is a redshift evolution, because of the systematic uncertainties \citep{Woo2006} and selection effects \citep{Lauer2007, Treu2007, Schulze2011, Schulze2014}.

At high-$z$, AGN is our only proxy for studying the BH mass assembly.
The luminosity functions (LFs) of AGN however remain uncertain. 
For example, BH mass function at $z=6$ has been inferred from optical AGN LFs in \citet{Willott2010a}. 
On the other hand, several works \citep{Wang2010, Volonteri2011, Fiore2012, Volonteri2016} have argued that there are large populations of obscured, accreting BHs at high-$z$. 
For instance, the BH mass density at $z=6$ from X-ray observations of AGN has been shown to be greater than that inferred from optical quasars by an order of magnitude or more \citep{Treister2011, Willott2011}. 
A luminosity dependent correction for the obscured fraction is proposed in \citet{Ueda2014} and the obscured fraction tends to increase with redshift up to $z\sim4$ \citep{Merloni2014, Vito2014, Buchner2015, Vito2018}. 
Regardless of the exact amount of the obscured AGNs, it is certain that a fraction of AGNs is obscured, and therefore missed by observations. 
As a result, quantities such as the BH mass function, BH mass density, and BH accretion rate density are still uncertain at high-$z$.

Here, we use the \textsc{BlueTides} simulation \citep{Feng2015} to make predictions for both the global BH mass properties and the scaling relations ($M_{\bullet}-M_{\star}$ and $M_{\bullet}-\sigma$ relations) from $z=8$ to $z=10$. 
\textsc{BlueTides} is a large-scale and high-resolution cosmological hydrodynamic simulation with $ 2 \times 7040^3 $ particles in a box of $400h^{-1}Mpc$ on a side, which includes improved prescriptions for star formation, BH accretion, and associated feedback processes. 
With such high resolution and large volume, we are able to study the scaling relations and the global properties of BH mass at high-$z$ for the first time. 
So far, various quantities measured in \textsc{BlueTides} have been shown to be in good agreement with all current observational constraints in the high-z universe such as UV luminosity functions \citep{Feng2016, Waters2016a, Waters2016b, Wilkins2017}, the first galaxies and the most massive quasars \citep{Feng2015, DiMatteo2017, Tenneti2018}, the Lyman continuum photon production efficiency \citep{Wilkins2016, Wilkins2017}, galaxy stellar mass functions \citep{Wilkins2018}, and angular clustering amplitude \citep{Bhowmick2017}.

The paper is organized as follows. 
In Section~\ref{sec:Methods}, we briefly describe the \textsc{BlueTides} simulation and several physics implementations. 
In Section~\ref{sec:GlobalQuantity}, we report BH mass properties: mass function, mass density, and accretion rate.  
In Section~\ref{sec:ScalingRelation}, we demonstrate the scaling relations between $M_{\bullet}$ and $M_{\star}$, and $\sigma$. 
In Section~\ref{sec:Discussion}, we study the selection effects for several galaxy properties on the scaling relations. 
In Section~\ref{sec:GrowthHistory}, we investigate the assembly history of how BHs evolove on $M_{\bullet}-M_{\star}$ and $M_{\bullet}-\sigma$ planes. 
In Section~\ref{sec:Conclusions}, we summarize the conclusion of the paper.

\section{Methods}
\label{sec:Methods}

\begin{table}
    \centering
    \caption{Numerical parameters for the \textsc{BlueTides} simulation.}
    \label{tab:ParaBT}
    \begin{tabular}{llll}
        \hline
        $h$ & 0.697 & $Boxsize$ & $400h^{-1}Mpc$ \\
        $\Omega_{\Lambda}$ & 0.7186 & $N_{particle}$ & $2 \times 7040^3$ \\
        $\Omega_{matter}$ & 0.2814 & $M_{DM}$ & $1.2 \times 10^7h^{-1}M_{\sun}$ \\
        $\Omega_{baryon}$ & 0.0464 & $M_{gas}$ & $2.36 \times 10^6h^{-1}M_{\sun}$ \\
        $\sigma_8$ & 0.820 & $\epsilon$ & $1.5h^{-1}kpc$ \\
        $n_s$ & 0.971 & $M_{\bullet,seed}$ & $5 \times 10^5h^{-1}M_{\sun}$ \\
        \hline
    \end{tabular}
\end{table}

\subsection{\textsc{BlueTides} hydrodynamic simulation}
\label{subsec:BlueTidesSimulation}
The \textsc{BlueTides} simulation has been carried out using the
Smoothed Particle Hydrodynamics code \textsc{MP-Gadget} on the Blue
Waters system at the National Center for Supercomputing Applications.
The hydrodynamics solver in \textsc{MP-Gadget} adopts the new
pressure-entropy formulation of smoothed particle hydrodynamics
\citep{Hopkins2013}.  This formulation avoids non-physical surface
tensions across density discontinuities.  \textsc{BlueTides} contains
$ 2 \times 7040^3 $ particles in a cube of $400 h^{-1} Mpc$ on a side
with a gravitational smoothing length $\epsilon=1.5h^{-1}kpc$.  The
dark matter and gas particles masses are
$M_{\rm DM}=1.2 \times 10^7 h^{-1}M_{\sun}$ and
$M_{\rm gas}=2.36 \times 10^6 h^{-1}M_{\sun}$, respectively.  The
cosmological parameters used were based on the Wilkinson Microwave
Anisotropy Probe nine years data \citep{Hinshaw2013} (see
Table~\ref{tab:ParaBT} for a brief summary of the parameters).  With
an unprecedented volume and resolution, \textsc{BlueTides} runs from
$z=99$ to $z=8$.  \textsc{BlueTides} contains approximately $200$
million star-forming galaxies, $160000$ of which have stellar mass
$>10^8M_{\sun}$, and $50$ thouthand BHs, $14000$ of which have BH mass
$>10^6M_{\sun}$ (the most massive BH's mass
$\sim 4 \times 10^8M_{\sun}$).  A full description of
\textsc{BlueTides} simulation can be found in \citet{Feng2016}.

\subsection{Sub-grid physics and BH model}
\label{subsec:BHmodel}
A number of physical processes are modeled via sub-grid prescriptions for galaxy formation in \textsc{BlueTides}. 
Below we list the main features of the sub-grid models:
\begin{itemize}
\item Star formation based on a multiphase star formation model
  \citep{Springel2003} with modifications following
  \citet{Vogelsberger2013}.
\item Gas cooling through radiative processes \citep{Katz1996} and metal cooling \citep{Vogelsberger2014}. 
\item Formation of molecular hydrogen and its effects on star formation \citep{Krumholz2011}. 
\item Type II supernovae wind feedback (the model used in {\it Illustris} \citep{Nelson2015}). 
\item A model of 'patchy' reionization \citep{Battaglia2013} yielding a mean reionization redshift $z\sim10$ \citep{Hinshaw2013}, and incorporating the UV background estimated by \citet{FaucherGiguere2009}; 
\item Black growth and AGN feedback. BHs grow in mass by gas accretion and by merging with other BHs.
\end{itemize}

We model BH growth and AGN feedback in the same way as in the {\it
  MassiveBlack I} \& {\it II} simulations, using the SMBH model
developed in \citet{DiMatteo2005} with modifications consistent with
{\it Illustris}.  BHs are seeded with an initial seed mass of
$M_{\rm \bullet,seed} = 5\times10^{5}h^{-1}M_{\sun}$ (commensurate with
the resolution of the simulation) in halos more massive than
$5\times10^{10}h^{-1}M_{\sun}$ while their feedback energy is
deposited in a sphere of twice the radius of the SPH smoothing kernel
of the black hole.  Gas accretion proceeds via
$\dot{M}_{\bullet} = \frac{ 4 \pi \alpha G^2 M_{\bullet}^2 \rho }{
  \left( c_s^2 + v^2 \right)^{-3/2} } $
according to \citet{Hoyle1939, Bondi1944, Bondi1952}, where $\rho$ and
$c_s$ are the density and sound speed of the gas respectively,
$\alpha$ is a dimensionless parameter, and $v$ is the velocity of the
BH relative to the gas.  We allow for super-Eddington accretion but
limit the accretion rate to three times of the Eddington rate:
$\dot{M}_{\rm Edd} = \frac{ 4 \pi G M_{\bullet} m_p }{ \eta \sigma_T c }$,
where $m_p$ is the proton mass, $\sigma_T$ is the Thomson
cross-section, and $\eta$ is the radiative efficiency.  Fixing $\eta$
to $0.1$ according to \citet{Shakura1973} throughout the simulation,
the BH is assumed to radiate with a bolometric luminosity ($L_{\rm B}$)
proportional to the accretion rate ($\dot{M}_{\bullet}$) by
$L_B = \eta \dot{M}_{\bullet} c^2$.  The Eddington ratio is defined as
$\lambda_{\rm Edd} = \frac{ \dot{M_{\bullet}} }{ \dot{M}_{\rm Edd} }$.

\begin{figure}
        \includegraphics[width=\columnwidth]{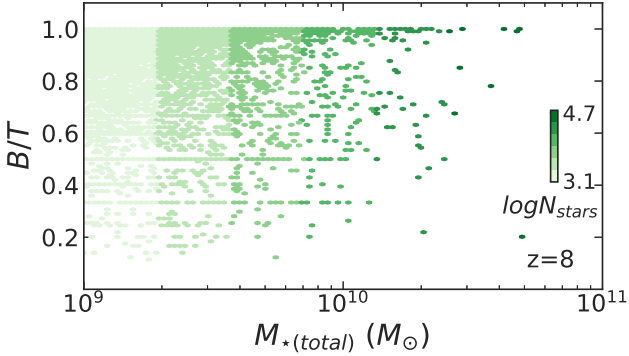}
        \caption{The relation between $B/T$  total stellar mass ($M_{\star}$) color coded according to number of star particles for each galaxy at $z=8$ in \textsc{BlueTides}.} 
        \label{fig:BtoT_Nstar}
\end{figure}

\subsection{Kinematic decomposition}
\label{subsec:kinematicdecomposition}
As $\sigma$ or $M_{\star}$ in observational studies of
$M_{\bullet}-\sigma$ or $M_{\bullet}-M_{\star}$ relations are often
measured from the bulge component of galaxies, we perform a kinematic
decomposition for the stellar particles of the galaxies in
\textsc{BlueTides} as in \citet{Feng2015}.  This allows us to
determine which stars are on planar circular orbits and which are
associated with a bulge, in each galaxy \citep{Vogelsberger2014,
  Tenneti2016}, providing kinematically classified disks and bulges,
and a disk to total ($D/T$) ratio for our galaxies.  We perform this
analysis following \citet{Abadi2003}: a circularity parameter is
defined for every star particle as $\kappa = j_z/j \left( E \right)$,
where $j_z$ is the specific angular momentum around a selected z-axis
and $j \left( E \right)$ is the possible maximum specific angular
momentum of the star with the specific binding energy $E$.  The star
particle with $\kappa>0.7$ is identified as a disk component according
to \citet{Vogelsberger2014} and \citet{Tenneti2016}.  Thus, the $D/T$
ratio for the stellar component of each galaxy is obtained, allowing
us to calculate the bulge stellar mass and the bulge velocity
dispersion for our galaxies.  In Figure~\ref{fig:BtoT_Nstar}, we show
the relation between $B/T=1-D/T$ (bulge to total ratio) and total
stellar mass ($M_{\star}$) color coded according to number of star
particles for each galaxy. According to the standard assumption $D/T <
0.3$ is considered a bulge dominated galaxies \citep{Feng2015, Tenneti2016}. 
For galaxies with $M_{\star} > 10^9 M_{\sun}$,
the number of star particles is higher than $1000$. We require this
minimum number of star particles to have a reliable kinematic
decomposion. For this reason, for the rest of the analysis we will
only consider objects with $M_{\star} > 10^9 M_{\sun}$.

\section{The global property of BH mass}
\label{sec:GlobalQuantity}
We begin by investigating the global properties of BH mass ($M_{\bullet}$) from $z=8$ to $z=12$ in \textsc{BlueTides}. 
We choose a BH population with $M_{\bullet} > 1.5 \times 10^6 M_{\sun}$ which is roughly twice the BH seed mass ( $M_{\rm \bullet ,seed} = 7.2\times10^5M_{\sun}$), in order to minimize any possible influence of the seeding prescription on our analysis. 

\begin{figure*}
\begin{subfigure}[t]{\columnwidth}
    \includegraphics[width=\columnwidth]{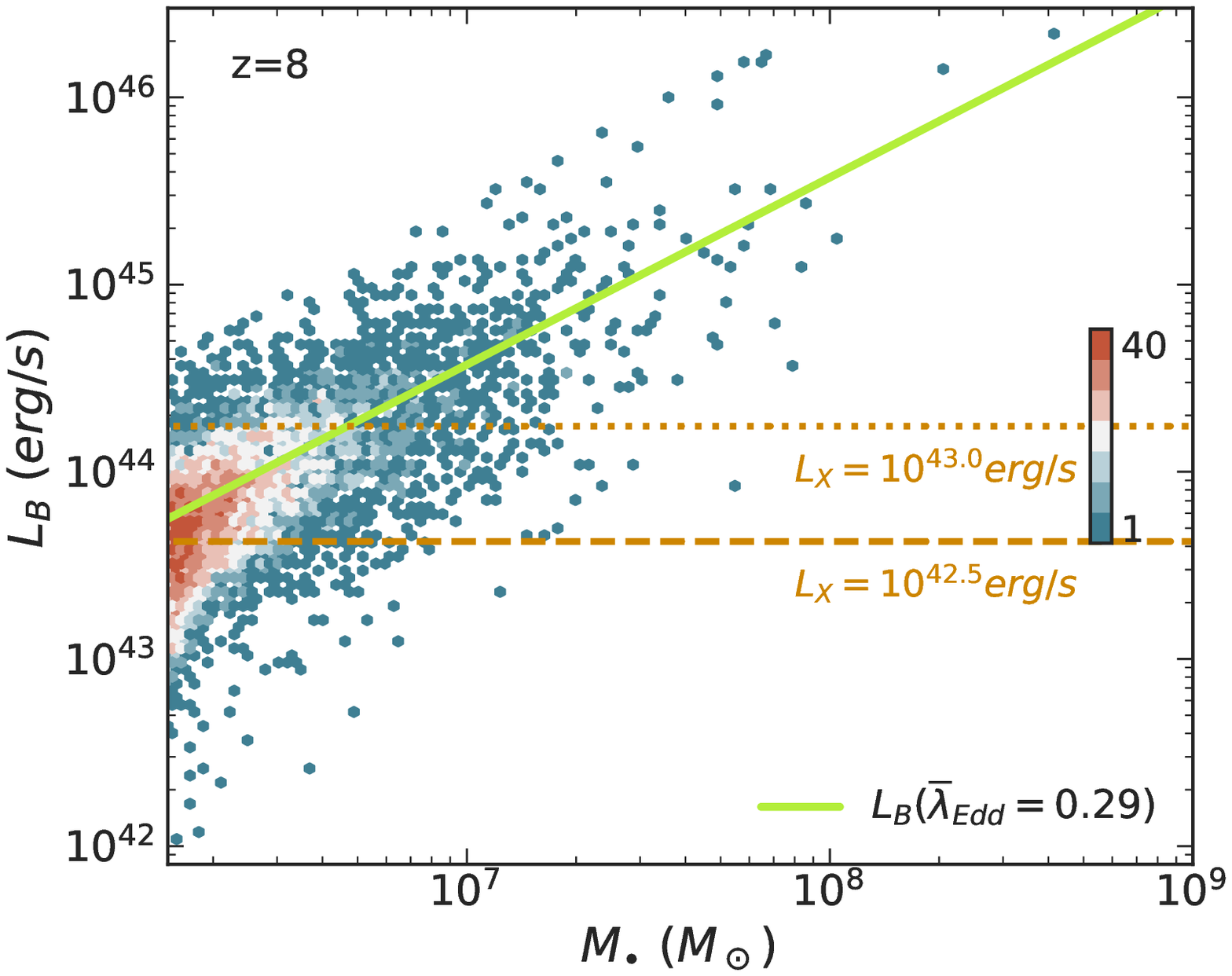}
    \end{subfigure} 
    \begin{subfigure}[t]{\columnwidth}
    \includegraphics[width=\columnwidth]{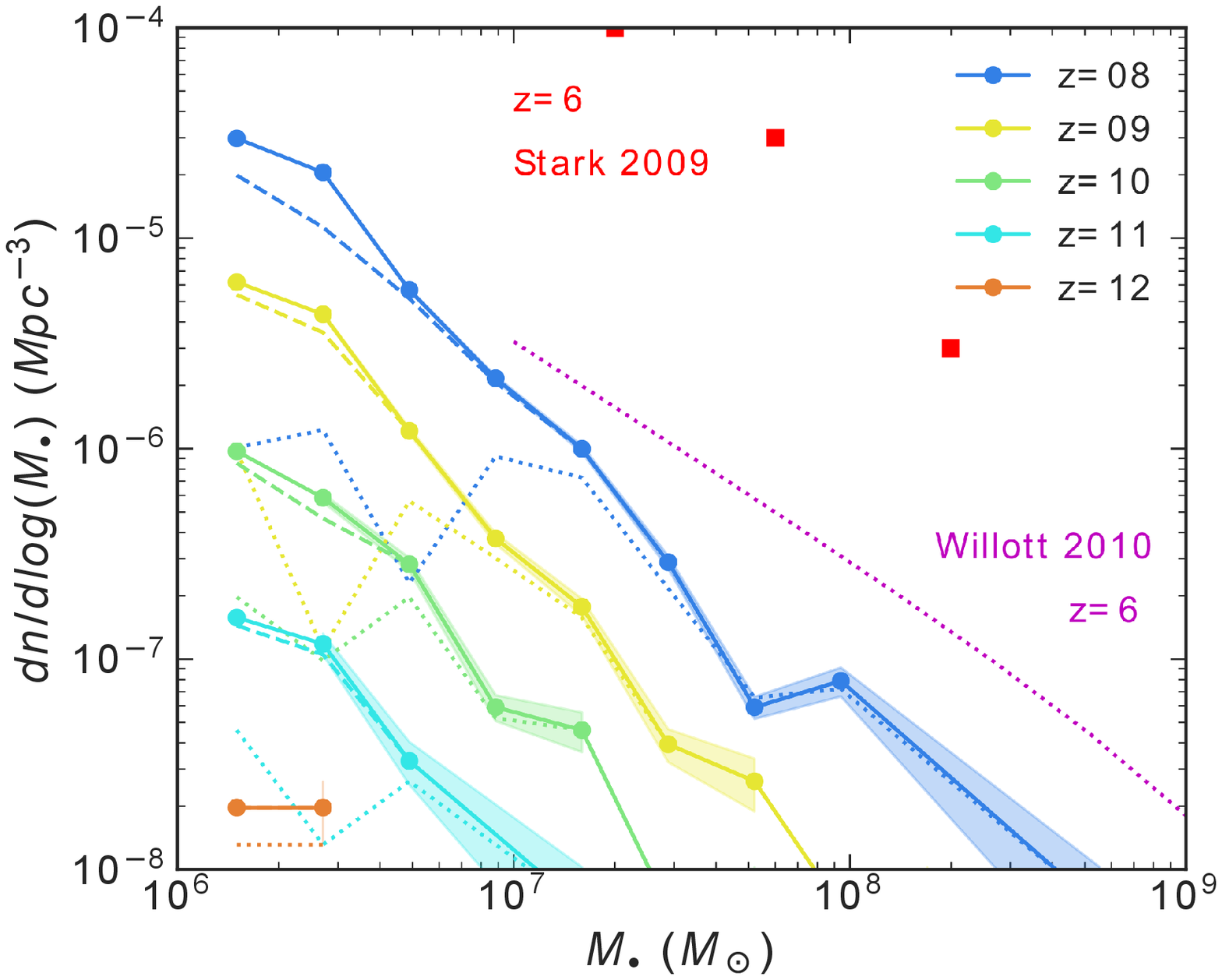}
    \end{subfigure} 
    \caption{Left panel: the relation between BH bolometric luminosity ($L_{\rm B}$) and BH mass ($M_{\bullet}$) at $z=8$ in \textsc{BlueTides} color coded according to the number of galaxies. The green line shows $L_{\rm B}$ with the mean Eddington ratio of our all BH population ($L_{\rm \overline{\lambda}_{\rm Edd} = 0.29}$). The brown dashed and dotted lines show the X-ray luminosity $L_X=10^{42.5}$ and $10^{43}erg/s$ respectively according to the bolometric correction from \citet{Marconi2004}. Right panel: BH mass functions in \textsc{BlueTides} at $z=8\sim12$ (the solid curves). The dashed and dotted curves show the BH mass functions with thresholds of $L_{\rm X}=10^{42.5}$ and $10^{43}erg/s$ respectively at the corresponding redshift. Also, the results at $z=6$ in \citet{Willott2010a} and \citet{Volonteri2011} are shown.}
    \label{fig:MassFunction}
\end{figure*}

\subsection{BH mass function and bolometric luminosity}
\label{sec:BHMF}
We first look at the bolometric luminosity ($L_{\rm B}$) of BH population in \textsc{BlueTides} at $z=8$ in the left panel in Figure~\ref{fig:MassFunction}. 
The brown dashed and dotted lines are X-ray luminosity $L_{\rm X} = 10^{42.5}$ and $10^{43} erg/s$, which are calculated by the bolometric correction in \citet{Marconi2004}. 
These two values will be used as thresholds when studying other global properties of BH mass in this section. 
Statistically, there are more than $76$ and $16$ percent of our BHs with $L_{\rm X} > 10^{42.5}$ and $ 10^{43} erg/s $ respectively.
This indicates that the global quantities of BH mass is sensitive to $L_{\rm X}$ when measured from X-ray survey in observation. 
In addition, $L_{\rm B}$ with the mean Eddington ratio in our BH population ($\overline{\lambda}_{\rm Edd} = 0.29$) is shown (the green solid line).

The right panel in Figure~\ref{fig:MassFunction} shows BH mass functions (BHMFs) in \textsc{BlueTides} from $z=8$ to $z=12$ (the solid curves), as well as the ones with thresholds $L_{\rm X} > 10^{42.5}$ and $ 10^{43} erg/s $ (the dashed and dotted curves respectively). 
We also show the BHMFs inferred from optical quasars at $z=6$ in \citet{Willott2010a} (W10 hereafter; the purple dotted curve) and the theoretical prediction combined with observed Lyman break galaxy population \citep{Stark2009, Volonteri2011} (the red squares). 
The slope of BHMFs in \textsc{BlueTides} are generally steeper than the one in W10 but similar to theoretical predictions (see \citet{Volonteri2011} for more detail and comparison),  particularly at the low mass end (which is currently unconstrained at these redshifts). 
In addition, the normalization of the BHMFs in \textsc{BlueTides} suggests that there is a larger BH population than those which have been observed from optical quasars, consistent with the claim that there is a large population of obscured accreting BHs at high-$z$. 

\begin{figure*}
        \begin{subfigure}[t]{\columnwidth}
                \includegraphics[width=\columnwidth]{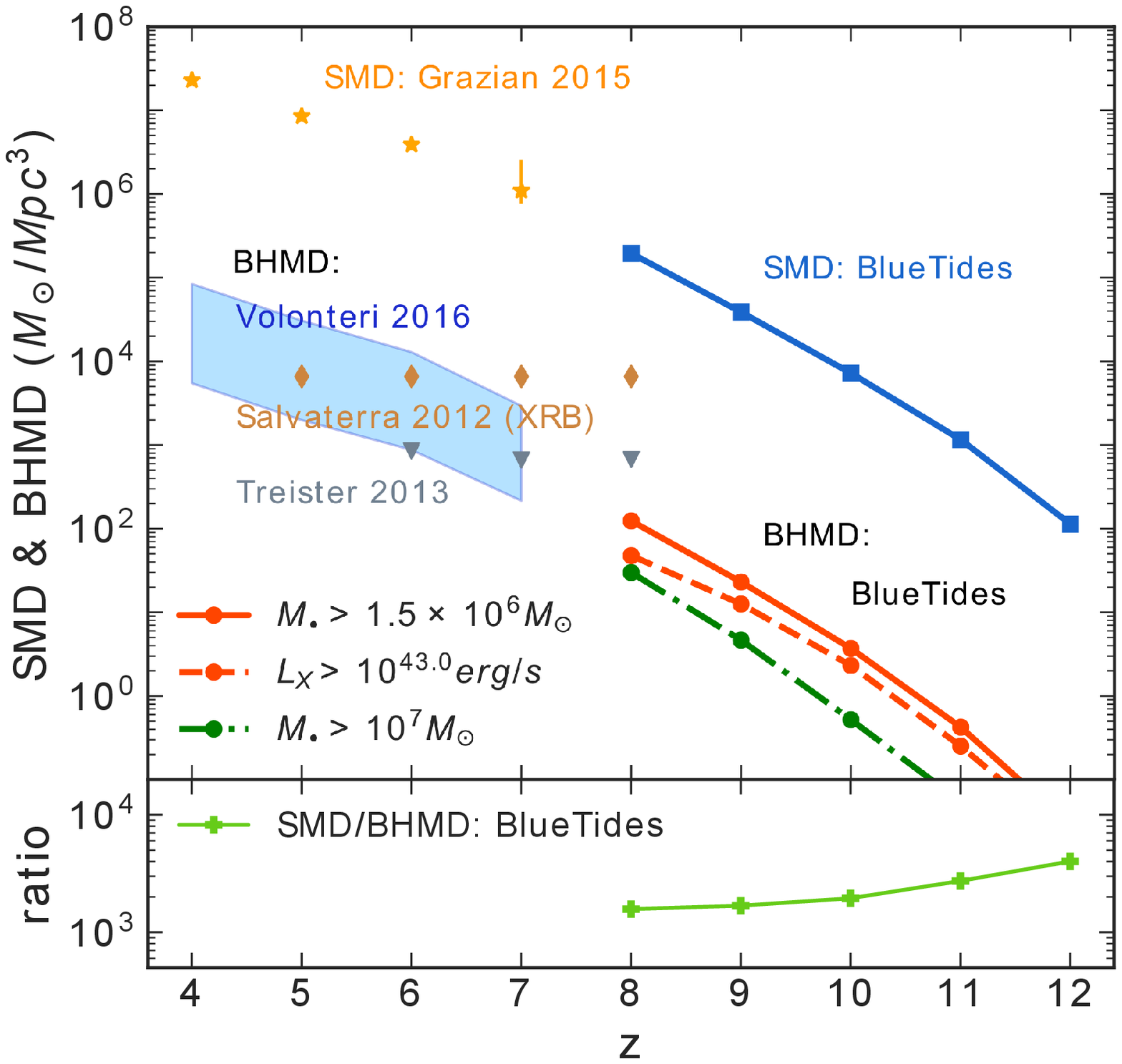}
        \end{subfigure} 
        \begin{subfigure}[t]{\columnwidth}
                \includegraphics[width=\columnwidth]{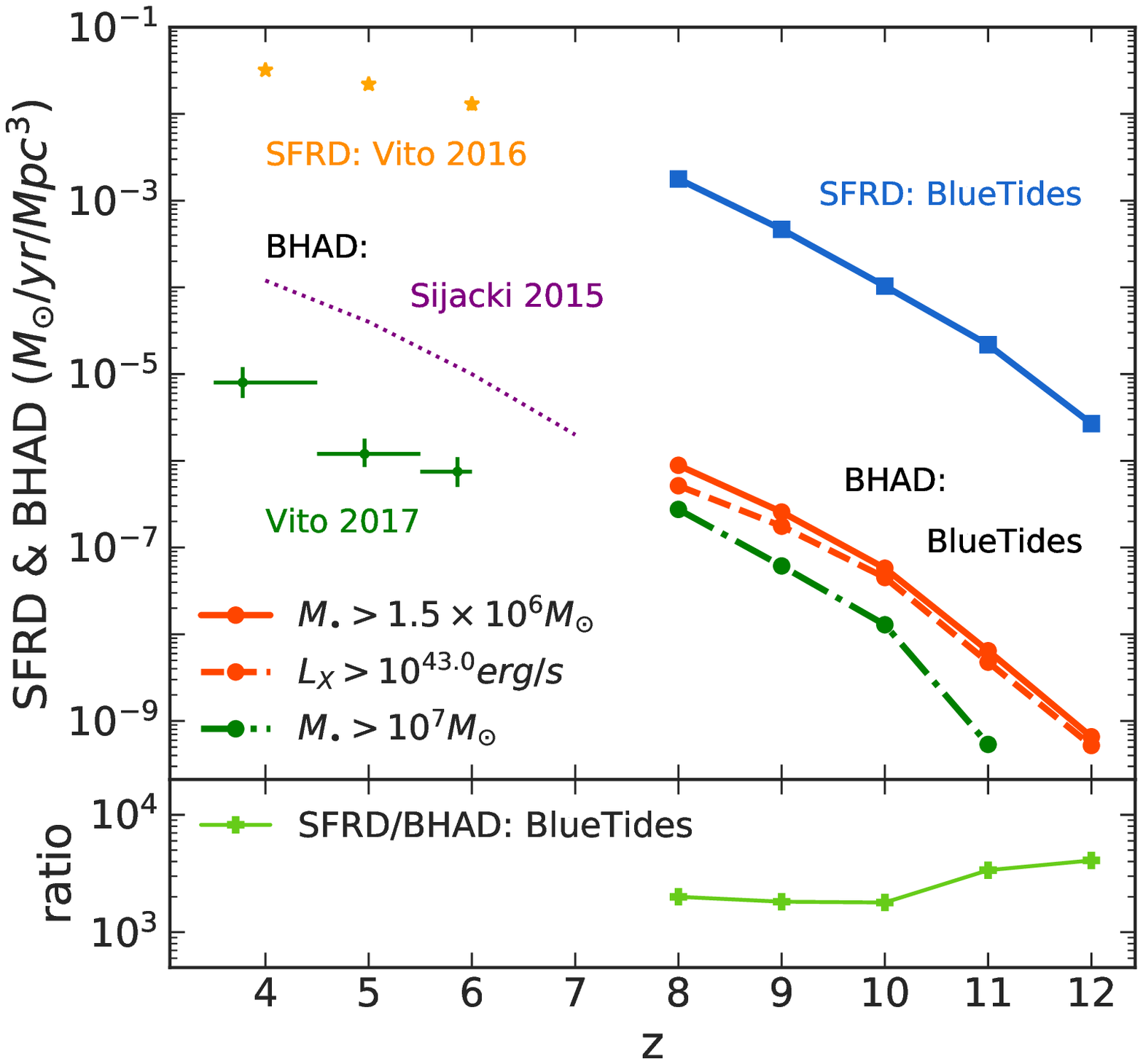}
        \end{subfigure} 
        \caption{Left panel: the stellar mass and BH mass density (SMD and BHMD) in\textsc{BlueTides}and their ratio SMD/BHMD (the green solid curve). For SMD, galaxies with $M_{\star}>10^8M_{\sun}$ are selected in simulation (the blue solid curve) and observation (\citet{Grazian2015}; the orange stars). For BHMD, galaxies with $M_{\bullet}>1.5\times10^6M_{\sun}$ (double of $M_{\rm \bullet ,seed}$), with $M_{\bullet}>10^7M_{\sun}$, and with $L_{\rm X}>10^{43}erg/s$ are shown (the red solid, olive dash-dotted, and red dash curves respectively). The blue shaded area is the result in \citet{Volonteri2016} and the brown diamonds and gray triangles are current upper limits from X-ray observations \citep{Salvaterra2012, Treister2013}. Right panel: the SFR and BH accretion rate density (SFRD and BHAD) in \textsc{BlueTides} and their ratio SFRD/BHAD (the green curve). The same thresholds are used as the left panel. Observational results in \citet{Vito2016} and \citet{Vito2018} for SFRD and BHAD are shown (the orange stars and green dots respectively), as well as the simulation prediction from \citet{Sijacki2015} (purple dotted curve; thresholds: $M_{\rm \bullet , seed} > 10^5 h^{-1} M_{\sun}$).}
        \label{fig:GSMDBHMD}
\end{figure*}

\subsection{BH mass density and stellar mass density}
\label{sec:MassDensity}
The left panel in Figure~\ref{fig:GSMDBHMD} shows BH and galaxy stellar mass density in \textsc{BlueTides} from $z=8$ to $z=12$ with observations from $z=4$ to $z=7$.
For stellar mass density (SMD), \textsc{BlueTides} (the solid blue curve) agrees with the trend from \citet{Grazian2015} (the orange stars). 
Galaxies with $M_{\star} > 10^8 M_{\sun}$ are selected for both cases. 
For BH mass density (BHMD), we report the results with two different $M_{\bullet}$ thresholds: $ M_{\bullet} > 1.5 \times 10^6 M_{\sun} $ ( double of $M_{\rm \bullet ,seed}$ in BlueTides) and $M_{\bullet} > 10^7M_{\sun}$ ) and with $L_{\rm X} > 10^{43 erg/s}$ (the red solid, olive dash-dotted, and red dash curves respectively) to show the influence from $M_{\bullet}$ and $L_{\rm X}$ thresholds. 
Current upper limits from X-ray observation are also presented: \citet{Salvaterra2012} (cosmic X-ray background (XRB) ) and \citet{Treister2013} (the brown diamonds and gray triangles respectively). 
BHMD in \textsc{BlueTides} complies with those upper limits (at least at $z=8$), pointing that our BH mass function is just steeper than the one in W10 but still within the upper limits from X-ray observation. 
In addition, results in \citet{Volonteri2016} (the blue shade) are also included to support our BHMF, arguing that the integrated BH density depends on the $M_{\bullet}-M_{\star}$ relation.

It has been discussed that the stellar mass density exceeds the BH mass density roughly by a factor of $10^3$ for low-$z$. 
To understand the ratio of these two quantities at higher redshifts, we show the ratio by normalizing SMD to BHMD in the left panel in Figure~\ref{fig:GSMDBHMD} (the green solid curve). 
Overall, SMD grows more rapidly than the BHMD at early times. 
Parameterizing the ratio by an evolutionary factor $(1+z)^\alpha$, we find that $ SMD/BHMD = 1.3 (1+z)^{3.1} $. 

\subsection{ BH accretion rate density and SFR density}
\label{sec:SFRD_BHARD}
After BHMD and SMD, we investigate their assembly rate: the BH accretion rate density and the SFR density. 
The right panel in Figure~\ref{fig:GSMDBHMD} shows the BH accretion rate density (BHAD) again with $M_{\bullet}$ and $L_{\rm X}$ thresholds $ M_{\bullet} > 1.5 \times 10^6 M_{\sun} $, $M_{\bullet} > 10^7M_{\sun}$, and $L_{\rm X}>10^{43} erg/s$ (the red solid, olive dash-dotted, and red dash curves respectively) and the SFR density (SFRD; the blue solid curve) in \textsc{BlueTides} from $z=8$ to $z=12$. 
For SFRD, we show observational result in \citet{Vito2016} (the orange stars), and on the other hand for BHAD, we not only show results from observation (\citet{Vito2018}; the green dots) but also from simulation (\citet{Sijacki2015}; the dotted purple curve) because it has been noticed that the prediction from simulation tends to be higher than current observation by more than an order of magnitude. 
It is possibly due to the difficulty of observing AGNs from deep X-ray surveys.

Similar to Section~\ref{sec:MassDensity}, we compare these two quantities by their ratio via normalizing SFRD to BHAD (the green curve). 
The ratio of the SFRD and the BHAD increases as $z$ increases and the order of which (ranging from $10^3$ to $10^4$) is close to the order of the ratio of the BHMD and the SMD in our simulation. 
Again, we fit the ratio as a function of $(1+z)^{\alpha}$: $ SFRD/BHARD = 3.4 (1+z)^{2.7}$. 

\section{The Scaling Relations}
\label{sec:ScalingRelation}

\begin{figure}
        \begin{subfigure}[t]{\columnwidth}
                \includegraphics[width=\columnwidth]{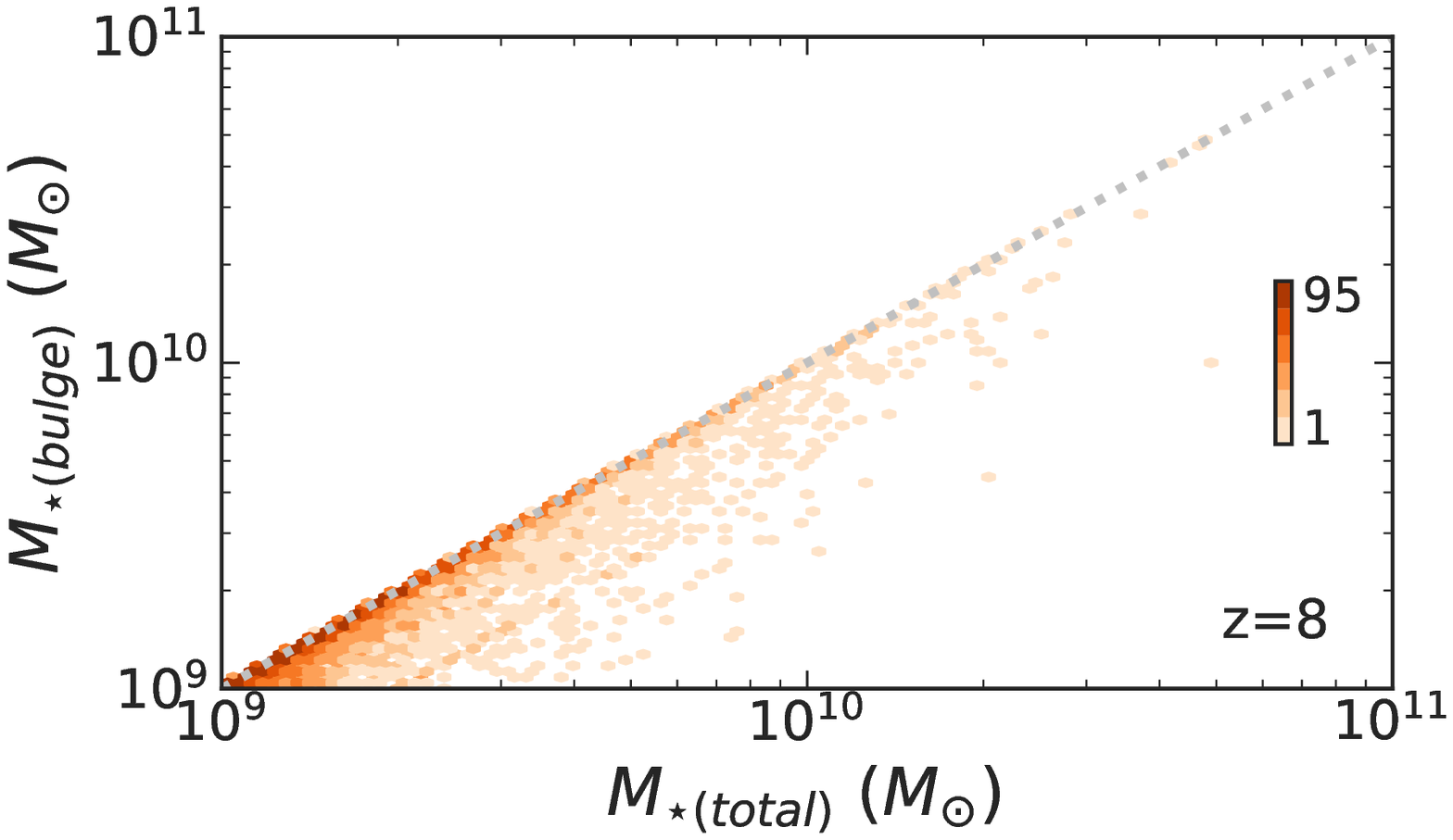}
        \end{subfigure} 
        \begin{subfigure}[t]{\columnwidth}
                \includegraphics[width=\columnwidth]{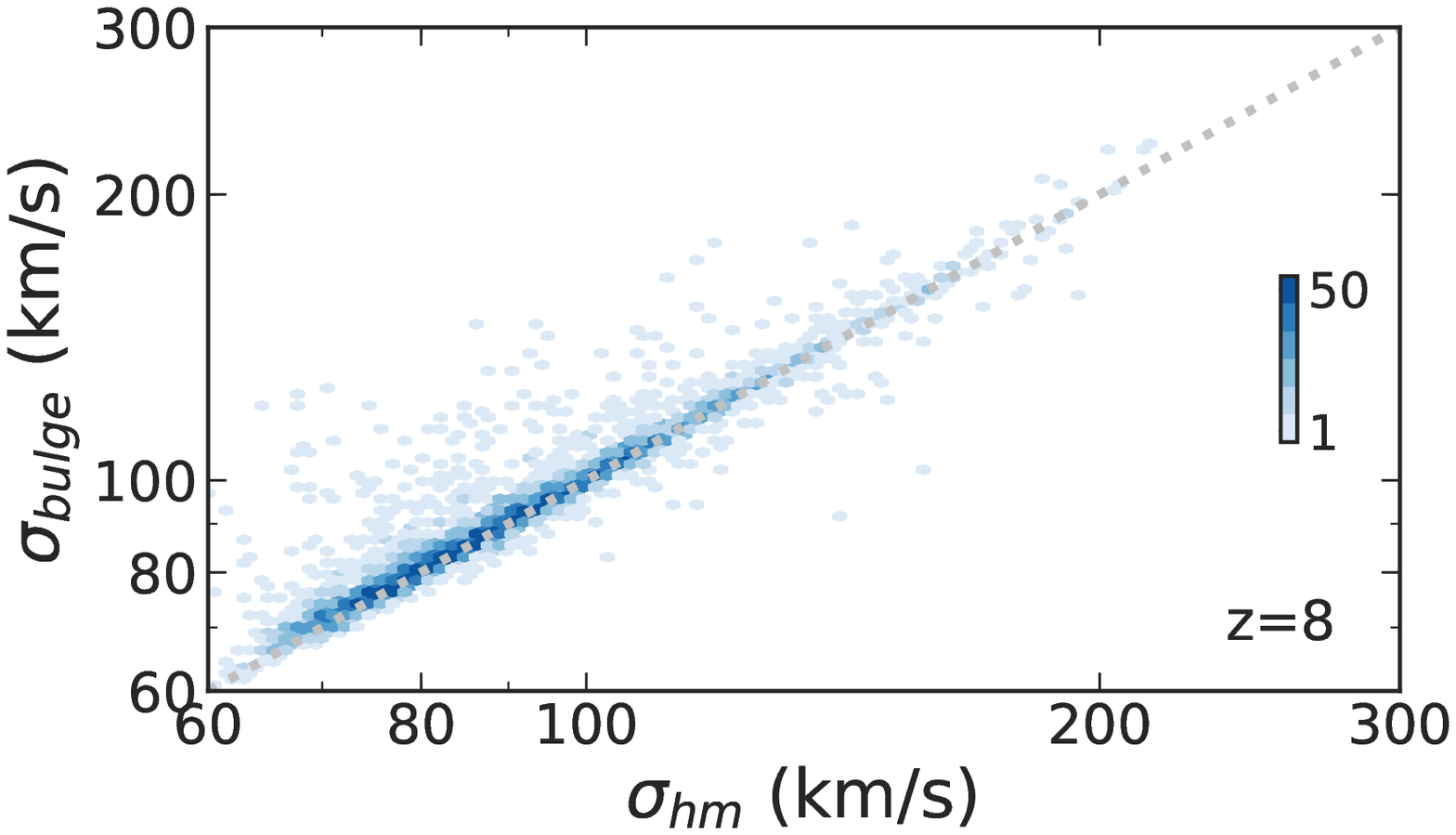}
        \end{subfigure} 
        \caption{The top and bottom panels show $M_{\star {\rm ,bulge} }$ versus $M_{\star {\rm ,total} }$ and $\sigma_{\rm bulge}$ versus $\sigma_{\rm hm}$ respectively, color coded by the number of galaxies at $z=8$ in \textsc{BlueTides}.} 
        \label{fig:ComparisonProxy}
\end{figure}

\subsection{Measuring $M_{\star}$ and $\sigma$}
\label{sec:Proxy}
The scaling relations between BH mass and their host galaxy properties have been measured by total stellar mass ($M_{\star  {\rm ,total} }$) or bulge stellar mass ($M_{\star {\rm ,bulge} }$) for $M_{\bullet}-M_\star$ relation, and by velocity dispersion of bulge stars for $M_{\bullet} -\sigma$ relation. 
In simulations, total or half stellar mass is available as a proxy for $M_{\star {\rm ,bulge} }$. 
A proxy for $\sigma$ often used is the velocity dispersion within half-light (or mass) radius (as for example in \citet{Sijacki2015} and \citet{DeGraf2015}). 
With the dynamical disk-bulge decomposition (see Section~\ref{subsec:kinematicdecomposition}) for the stellar components of galaxies in \textsc{BlueTides}, we directly have $M_{\star {\rm ,bulge} }$ and $\sigma_{\rm bulge}$ in our galaxies.

The top panel in Figure~\ref{fig:ComparisonProxy} shows the comparison between the $M_{\star {\rm ,bulge} }$ and $M_{\star  {\rm ,total} }$ color coded according to the number of galaxies. 
We find that about $82 \%$ of objects are bulge-dominated with $D/T < 0.3$. 
The bottom panel in Figure~\ref{fig:ComparisonProxy} shows the comparison between $\sigma_{\rm bulge}$ and $\sigma_{\rm hm}$ color coded according to the number of galaxies. 
Again, there is certainly a strong correlation between the two but an increased scatter above the one-to-one relation for a small number of objects, for which we have larger values of $\sigma_{\rm bulge}$ for a given $\sigma_{\rm hm}$. 
In particular, we find that over $94 \%$ and $97 \%$ of our galaxies have the difference between $\sigma_{\rm bulge}$ and $\sigma_{\rm hm}$ less than $10 \%$ and $20 \%$ respectively. 
We shall see in the next section how the detailed dynamical decomposition and the resulting $\sigma_{\rm bulge}$ and $M_{\star {\rm ,bulge} }$ impact the scaling relations. 

\begin{figure*}
    \includegraphics[width=1.8\columnwidth]{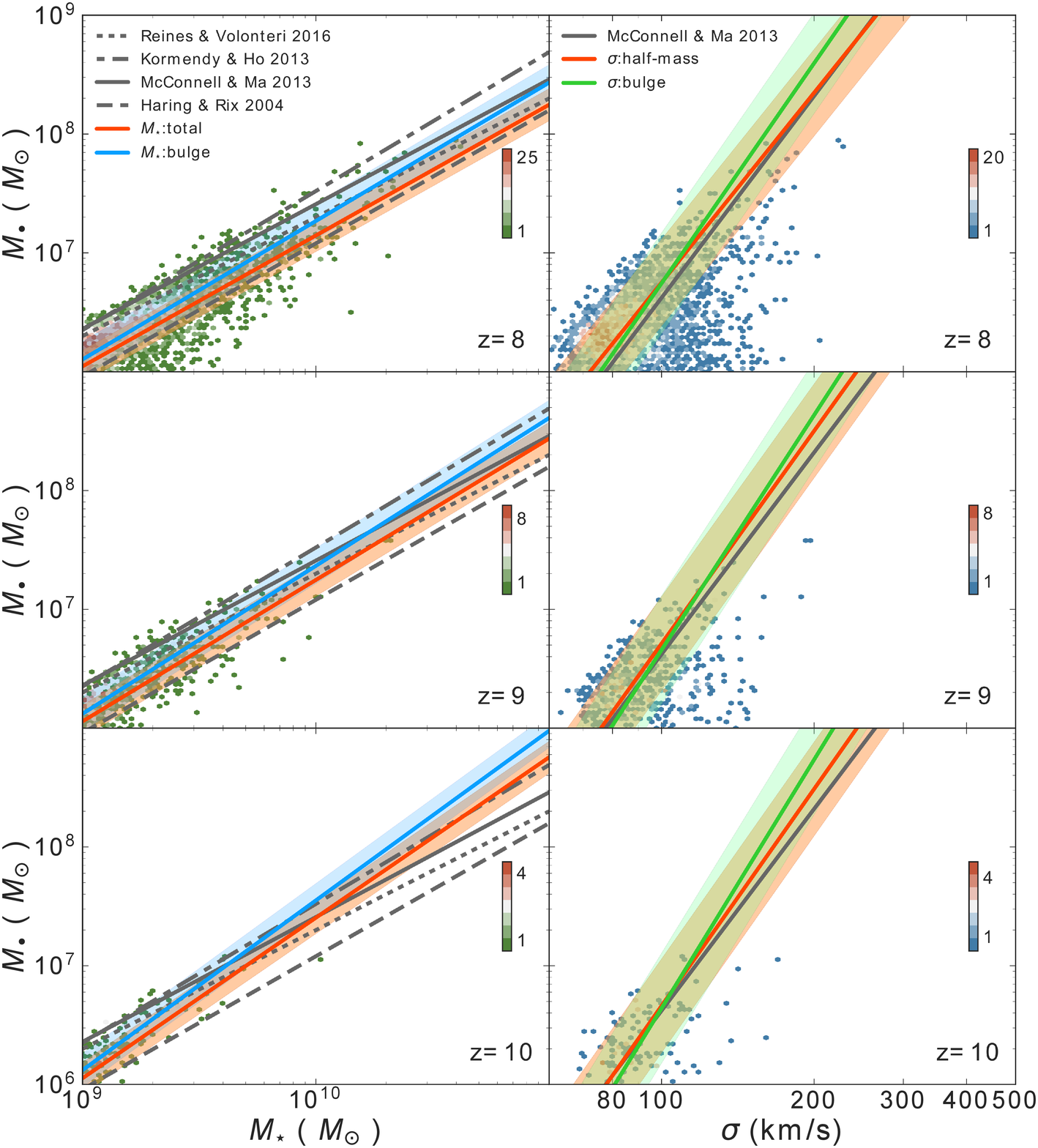}
    \caption{The scaling relations at $z=8$, $9$, and $10$ in \textsc{BlueTides} color coded according to the number of galaxies. Left panels: the $M_{\bullet}-M_{\star}$ relations with $M_{\star {\rm ,bulge} }$ for the data points. The red and blue lines show the best-fitting relation using $M_{\star {\rm ,total} }$ and $M_{\star {\rm ,bulge} }$ respectively while the gray lines show the observations \citep{Haring2004, McConnell2013, Kormendy2013, Volonteri2016}. Right panels: the $M_{\bullet}-\sigma$ relations with $\sigma_{\rm bulge}$ for the data points. The red and green lines show the best-fitting relation with $\sigma_{\rm hm}$ and $\sigma_{\rm bulge}$ respectively while the gray lines show the observations in \citet{McConnell2013}. The shaded area shows the standard deviation of residuals.}
    \label{fig:ScalingRelation}
\end{figure*}

\begin{table*}
    \centering
    \caption{The fitting coefficients $\alpha$ and $\beta$ (normalization and slope) of equation~(\ref{eq:FitForm}), the total number of data points $N$, and the standard deviation of residuals $\epsilon$ of the scaling relations at each redshift.}
    \label{tab:FitCoeffMain}
    \begin{tabular}{lcccccccccccccr} 
        \hline
        & & \multicolumn{3}{c}{$M_{\bullet}-M_{\star {\rm ,total} }$} &\multicolumn{3}{c}{$M_{\bullet}-\sigma_{\rm hm}$} & & \multicolumn{3}{c}{$M_{\bullet}-M_{\star {\rm ,bulge} }$} &\multicolumn{3}{c}{$M_{\bullet}-\sigma_{\rm bulge}$} \\
        \cline{3-8}\cline{10-15}
        $z$ & $N$ & $\alpha$ & $\beta$ & $\epsilon$ & $\alpha$ & $\beta$ & $\epsilon$ & & $\alpha$ & $\beta$ & $\epsilon$ & $\alpha$ & $\beta$ & $\epsilon$\\
        \hline
        8 & 8131 & $8.25_{\pm0.03}$ & $1.10_{\pm0.01}$ & 0.14 & $8.35_{\pm0.08}$ & $5.31_{\pm0.04}$ & 0.36 &  & $8.43_{\pm0.06}$ & $1.16_{\pm0.01}$ & 0.15 & $8.60_{\pm0.17}$ & $6.15_{\pm0.09}$ & 0.42\\
        9 & 1567 & $8.44_{\pm0.08}$ & $1.19_{\pm0.01}$ & 0.14 & $8.50_{\pm0.23}$ & $5.95_{\pm0.12}$ & 0.40 &  & $8.61_{\pm0.14}$ & $1.24_{\pm0.02}$ & 0.15 & $8.63_{\pm0.47}$ & $6.56_{\pm0.24}$ & 0.46\\
        10 & 269 & $8.76_{\pm0.20}$ & $1.35_{\pm0.02}$ & 0.13 & $8.49_{\pm0.54}$ & $6.06_{\pm0.28}$ & 0.40 &  & $8.98_{\pm0.37}$ & $1.43_{\pm0.04}$ & 0.14 & $8.73_{\pm1.24}$ & $6.95_{\pm0.63}$ & 0.46\\
        \hline        
    \end{tabular}
\end{table*}

\begin{figure}
        \begin{subfigure}[]{\columnwidth}
                \includegraphics[width=\columnwidth]{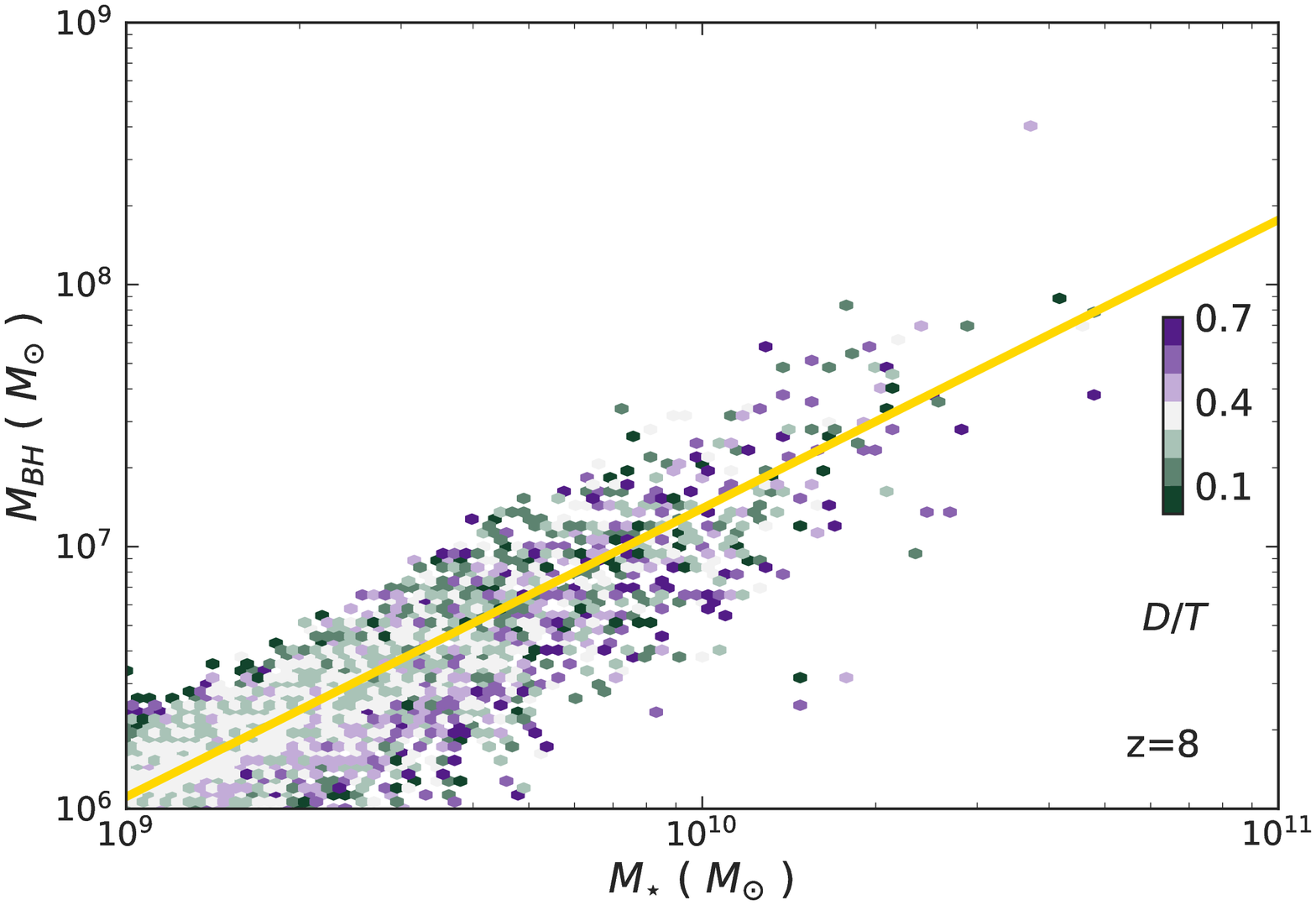}
        \end{subfigure} 
        \\
        \begin{subfigure}[]{\columnwidth}
                \includegraphics[width=\columnwidth]{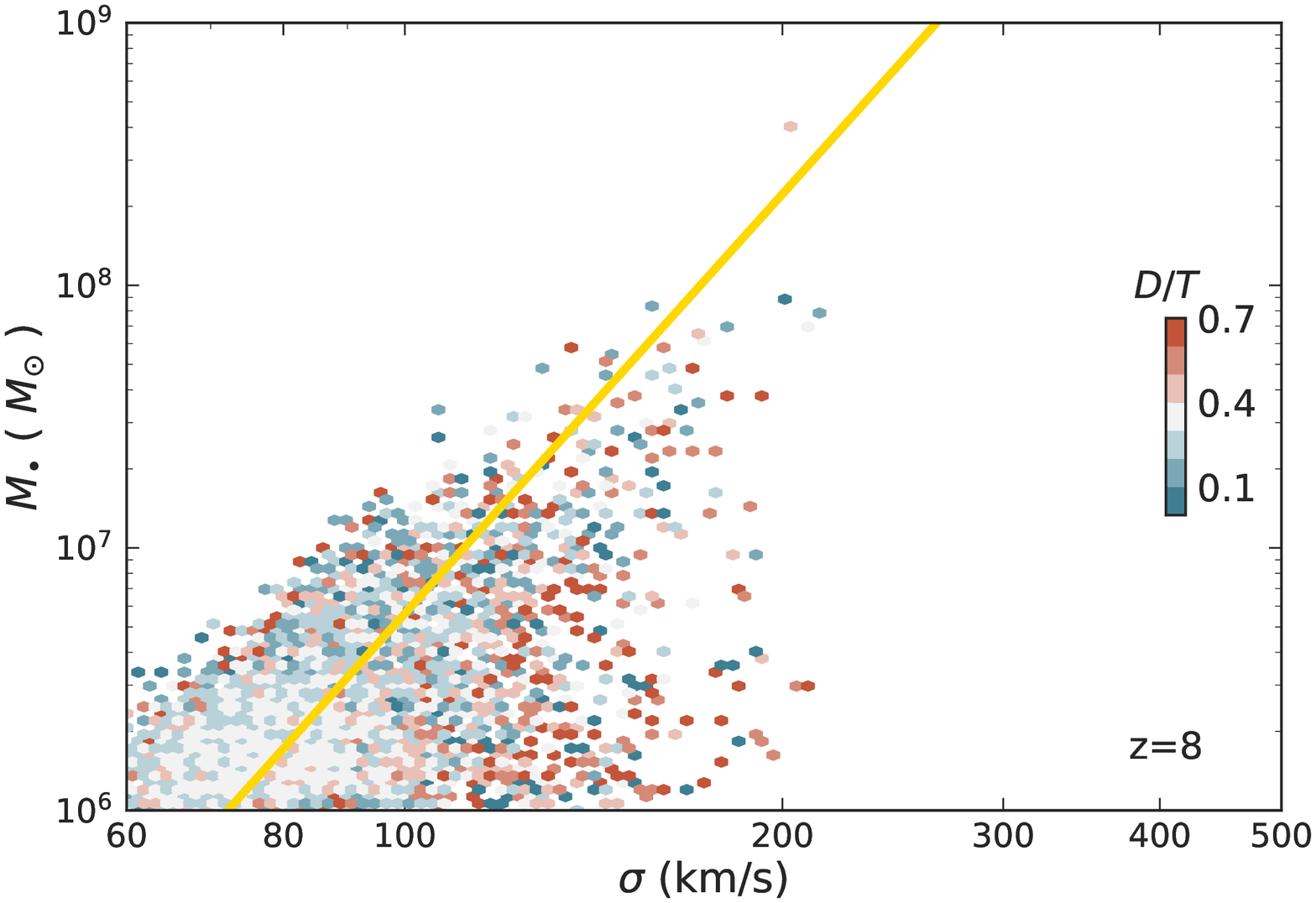}
        \end{subfigure} 
        \\
        \begin{subfigure}[]{\columnwidth}
                \includegraphics[width=\columnwidth]{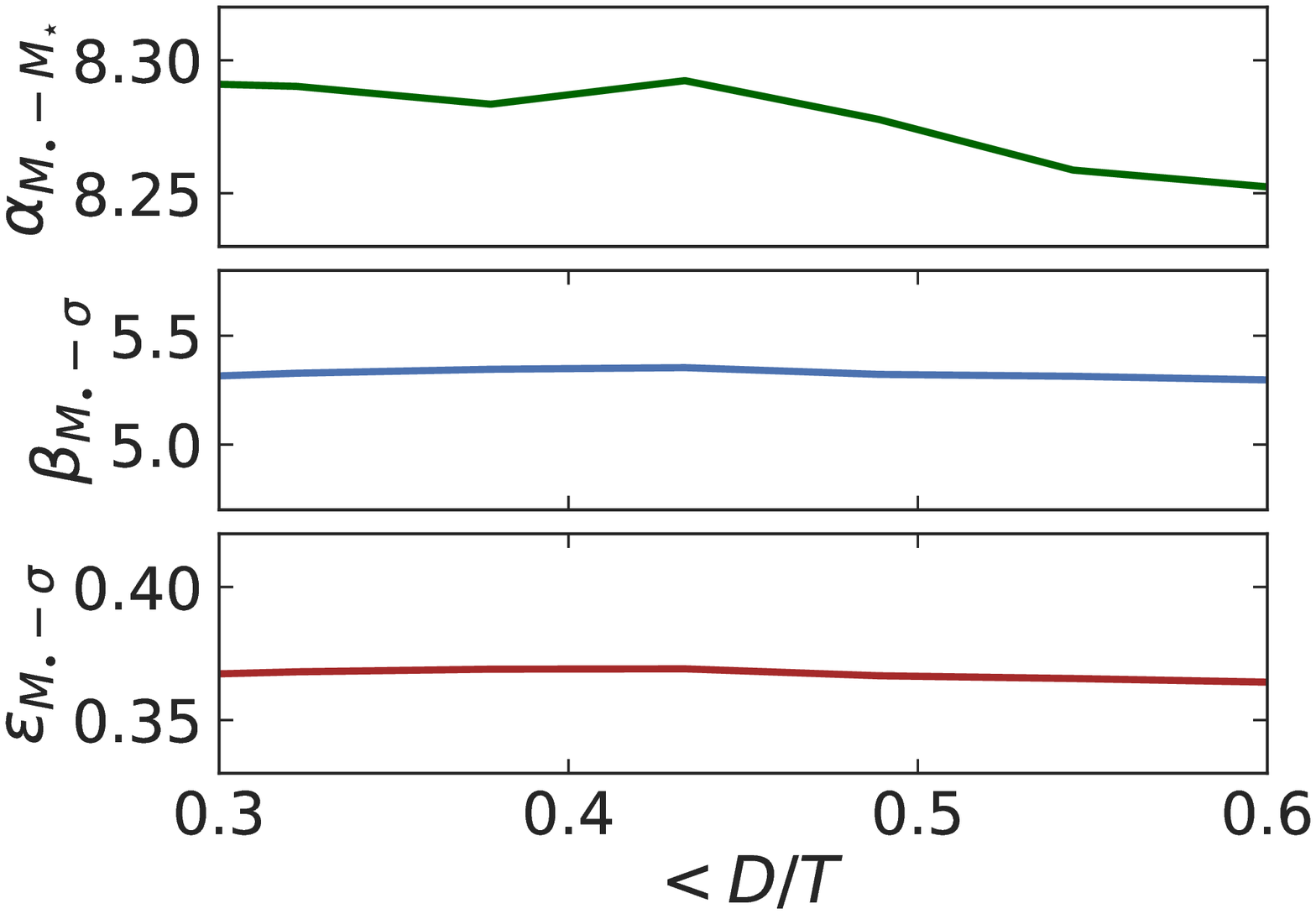}
        \end{subfigure} 
        \caption{Top and middle panels: the $M_{\bullet}-M_{\star}$ and $M_{\bullet}-\sigma$ relations color coded according to $D/T$ at $z=8$ in \textsc{BlueTides}. The yellow lines show the overall fits as the ones in Figure~\ref{fig:ScalingRelation}. Bottom panel: $\alpha$ of $M_{\bullet}-M_{\star}$ relation and $\beta$ and $\epsilon$ of $M_{\bullet}-\sigma$ relation as functions of different limiting $D/T$.} 
        \label{fig:bulgedominated}
\end{figure}

\subsection{$M_{\bullet}-M_\star$ and $M_{\bullet}-\sigma$ relation}
\label{subsec:scalingrelation}
Figure~\ref{fig:ScalingRelation} shows the $M_{\bullet}-M_{\star}$ relation and the $M_{\bullet}-\sigma$ relation at $z=8$, $9$, and $10$ in \textsc{BlueTides}, color coded according to the number of galaxies. 
Note that the data points are shown with $M_{\star {\rm ,bulge} }$ and $\sigma_{\rm bulge}$. 
We plot and fit only galaxies with $M_{\star} > 10^9M_{\sun}$ where a sufficient number of star particles are available to carry out the dynamical decomposition reliably. 
Both scaling relations in \textsc{BlueTides} are best-fitted by power laws as
\begin{equation}
	\log_{10} \left( M_{\bullet} \right) = \alpha + \beta \ \log_{10} \left( X \right),
    \label{eq:FitForm}
\end{equation}
where $M_{\bullet}$ is in units of $M_{\sun}$, and $X$ is $M_{\star}/10^{11}M_{\sun}$ or $\sigma/200kms^{-1}$.  
The fitting coefficients (normalization $\alpha$ and slope $\beta$) are summarized in Table~\ref{tab:FitCoeffMain}, including the total number of data points $N$ and the standard deviation of the residuals $\epsilon$.

The left panels in Figure~\ref{fig:ScalingRelation} show the $M_{\bullet}-M_{\star}$ relations. 
The red and blue lines show the best-fitting relation with $M_{\star {\rm ,total} }$ and $M_{\star {\rm ,bulge} }$ respectively while the gray lines show the observations: \citet{Haring2004}, \citet{McConnell2013} and \citet{Kormendy2013} with bulge stellar mass and elliptical samples while \citet{Volonteri2016} with total stellar mass. 
Our simulation provides the $M_{\bullet}-M_{\star}$ relation in the form of $\log_{10}(M_{\bullet}) = 8.25 + 1.10 \ \log_{10}(M_{\star}/10^{11}M_{\sun})$ with $M_{\star {\rm ,total} }$ at $z=8$, suggesting that the slopes are consistent with the observations but the normalizations are lower than most observations except for the one in \citet{Haring2004}. 
Both $\alpha$ and $\beta$ with $M_{\star {\rm ,total} }$ (the red lines) are lower than the ones with $M_{\star {\rm ,bulge} }$ (the blue lines) across all three redshifts, and both get steeper as $z$ is higher.  
The standard deviation of the residuals ($\epsilon$) is shown as the shaded area.

The right three panels in Figure~\ref{fig:ScalingRelation} show the $M_{\bullet}-\sigma$ relations.  
The red and green lines show the best-fitting relation with $\sigma_{\rm hm}$ and $\sigma_{\rm bulge}$ respectively while the gray lines show the observations in \citet{McConnell2013}.  
Our simulation provides the $M_{\bullet}-\sigma$ relation with $\sigma_{\rm hm}$ as $\log_{10}(M_{\bullet}) = 8.35 + 5.31 \ \log_{10}(\sigma/200kms^{-1})$ at $z=8$, which is consistent with the results of \citet{McConnell2013}.  
We note that both $\alpha$ and $\beta$ using $\sigma_{\rm hm}$ (the red lines) are lower by $\sim 3 \%$ and $\sim 10 \%$, respectively than the ones with $\sigma_{\rm bulge}$ (the blue lines) across all three redshifts, and both get steeper with increasing $z$ . 
Moreover, $M_{\bullet}-\sigma$ relations with $\sigma_{\rm bulge}$ are higher than local measurements. 
$\epsilon$ is shown as the shaded area and, more importantly, $M_{\bullet}-\sigma$ relation shows a larger scatter than the $M_{\bullet}-M_{\star}$ relation ($\epsilon \sim 0.4$ and $\epsilon \sim 0.1$ respectively) in our simulations. 
We will examine this in Section~\ref{sec:Discussion}.

For most observational results, the scaling relations are established with bulge-dominated galaxies. 
Here, we report how the relations change in our simulation with bulge-dominated galaxies ($D/T<0.3$). 
In the top two panels in Figure~\ref{fig:bulgedominated}, we show both relations color coded according to the $D/T$ ratio. 
In the bottom panels, we show $\alpha$ of $M_{\bullet}-M_{\star}$ relations and $\beta$ and $\epsilon$ of $M_{\bullet}-\sigma$ relations as functions of limiting $D/T$. 
We find that the relations hardly change even with different $D/T$, even for the bulge-dominated regime $D/T < 0.3$. 

\section{The slope and scatter in the scaling relations}
\label{sec:Discussion}
We have shown that the high-z relation is consistent with the locally
measured ones (both in slope and normalization).  However here we wish
to investigate possible selection effects and/or physical parameters
that may affect the slope and scatter in the scaling relations
(Figure~\ref{fig:ScalingRelation} and Table~\ref{tab:FitCoeffMain}).
In particular, we have found that there is a more significant scatter
in the $M_{\bullet}-\sigma$ relation (larger than in local
measurements) than the $M_{\bullet}-M_{\star}$ relation.  The
relatively large scatter in the $M_{\bullet}-\sigma$ relation appears
to be due to a significant amount of objects that lie below the main
relation: galaxies with relative high $\sigma$ compared to their
relative low $M_{\bullet}$.  Note that $M_{\star}$ denotes
$M_{\star {\rm ,total} }$ and $\sigma$ denotes $\sigma_{\rm hm}$
hereafter unless stated otherwise.

\begin{figure*}
        \begin{subfigure}[]{2\columnwidth}
                \includegraphics[width=\columnwidth]{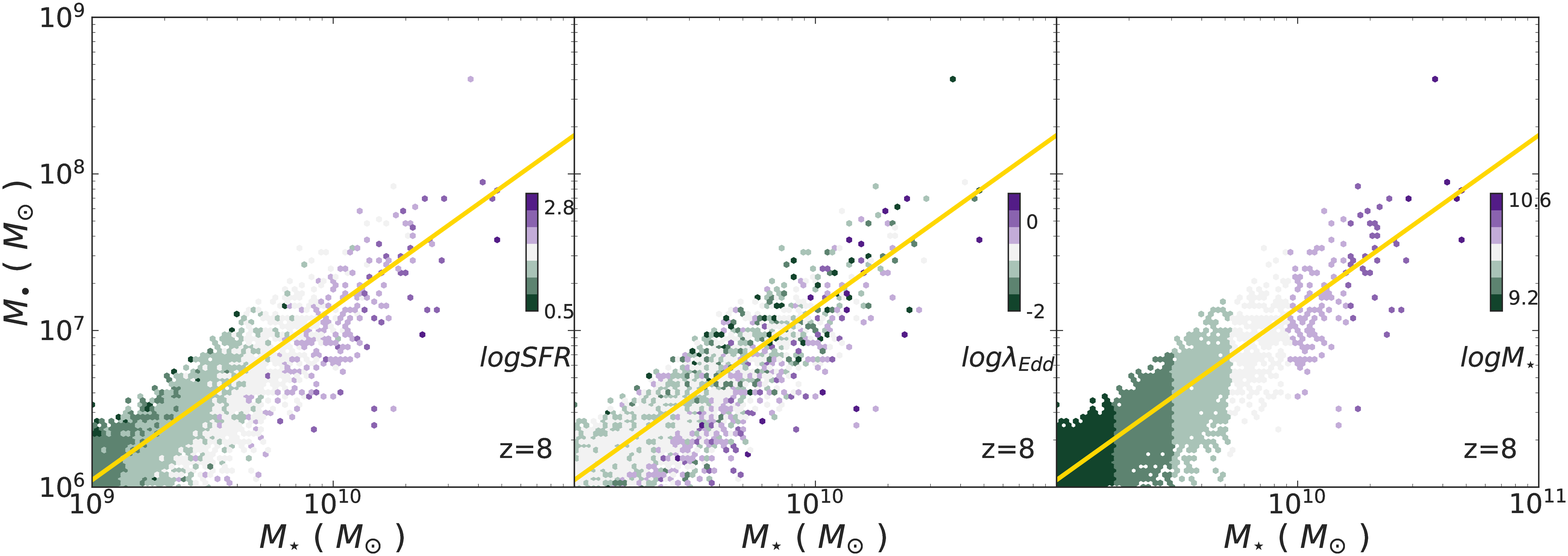}
        \end{subfigure} 
        \\
        \begin{subfigure}[]{2\columnwidth}
                \includegraphics[width=\columnwidth]{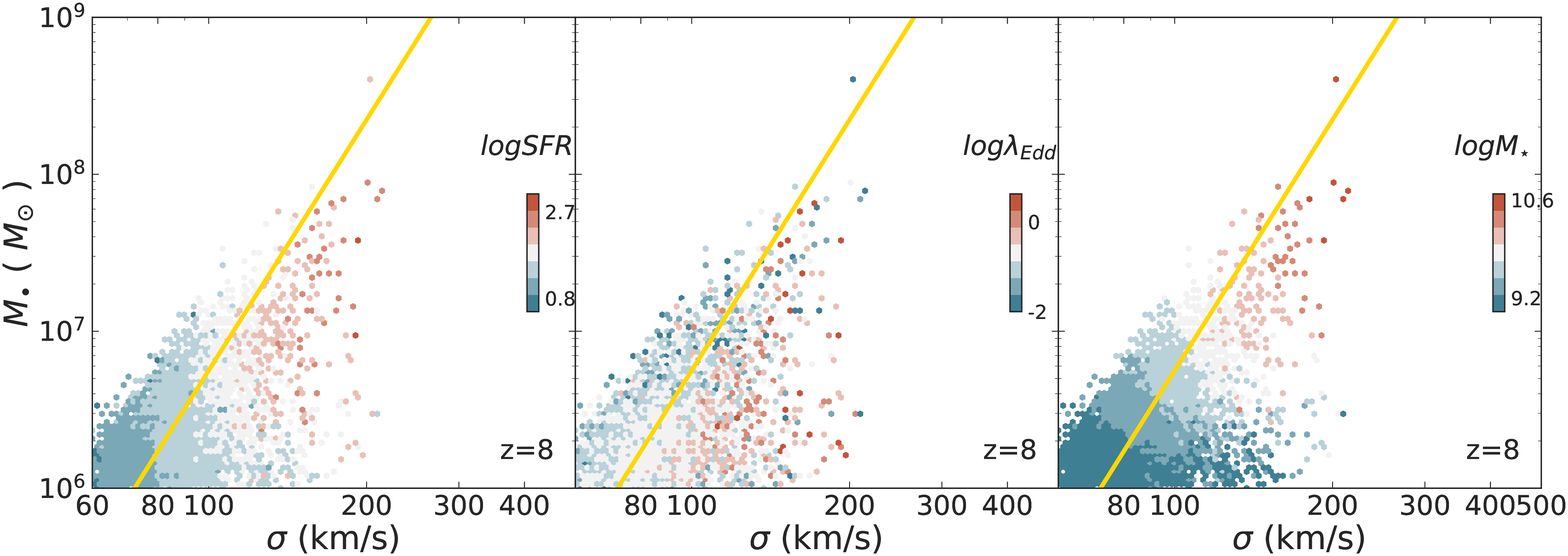}
        \end{subfigure} 
        \\
        \begin{subfigure}[]{2\columnwidth}
                \includegraphics[width=\columnwidth]{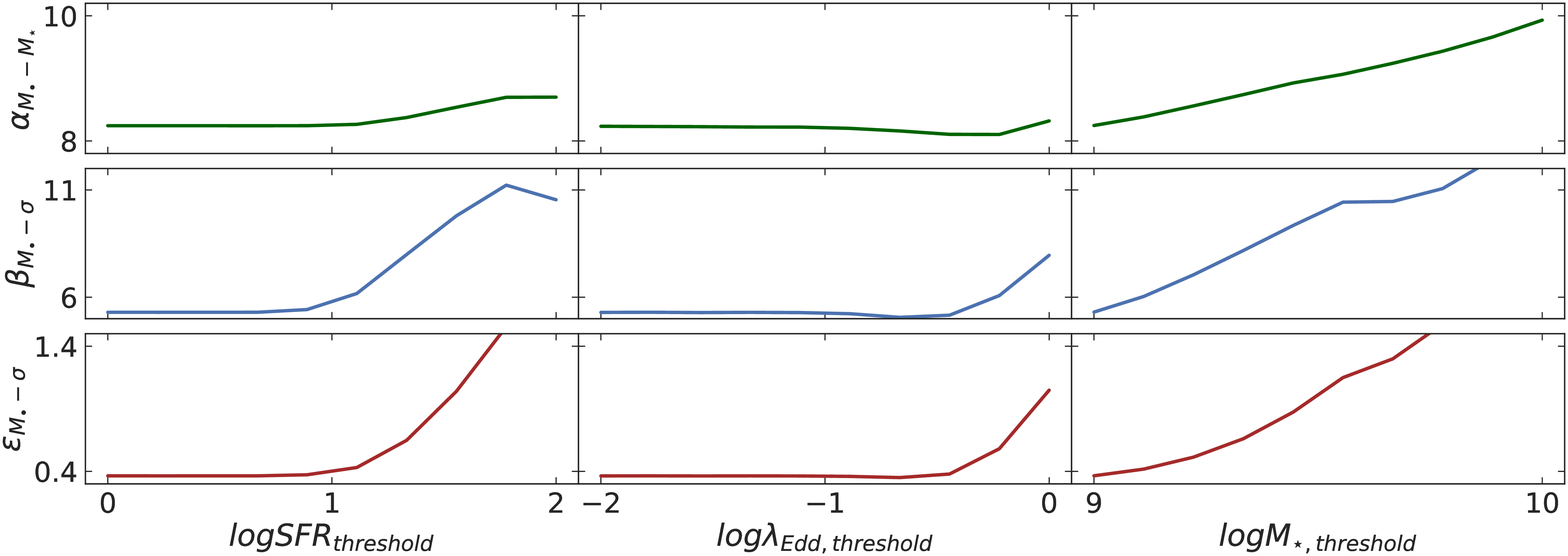}
        \end{subfigure} 
        \caption{Top and middle panels: the $M_{\bullet}-M_{\star}$ and $M_{\bullet}-\sigma$ relations color coded according to $SFR$, $\lambda_{\rm Edd}$, and $M_{\star}$ (from left to right respectively) at $z=8$ in \textsc{BlueTides}. The yellow lines show the overall fits as the ones in Figure~\ref{fig:ScalingRelation}. Bottom panel: $\alpha$ of $M_{\bullet}-M_{\star}$ relation and $\beta$ and $\epsilon$ of $M_{\bullet}-\sigma$ relation as functions of different thresholds of $SFR$, $\lambda_{\rm Edd}$, and $M_{\star}$ from left to right respectively.} 
        \label{fig:3Q}
\end{figure*}

\subsection{$SFR$, $\lambda_{\rm Edd}$, and $M_{\star}$ dependence}
\label{sec:dependence}
Here, we examine the dependency of the scaling relations at $z=8$ on
galaxy properties.  The top and middle panels in Figure~\ref{fig:3Q}
show the $M_{\bullet}-M_{\star}$ and $M_{\bullet}-\sigma$ relations
for threshold samples selected for (from the left to right) $SFR$ (a
proxy for the galaxy luminosity), stellar mass ($M_{\star}$), and BH
accretion rate, in units of Eddington (Eddington ratio
$\lambda_{\rm Edd}$; a proxy for the AGN luminosity), while the yellow
line is the best fit from all galaxies.  We wish to examine if the
above selection thresholds may have an influence on the slopes,
normalization and scatter in the relations.

The bottom panels of Figure~\ref{fig:3Q} show the variation in normalization $\alpha$ of the $M_{\bullet}-M_{\star}$ relation and in the slope $\beta$ and scatter $\epsilon$ of the $M_{\bullet}-\sigma$ relation as functions of thresholds of $SFR$, $\lambda_{\rm Edd}$, and $M_{\star}$.
We find that the normalization ($\alpha$) of $M_{\bullet}-M_{\star}$ relation increases slightly for higher thresholds of $SFR$ and $M_{\star}$ (while it is not sensitive to the Eddington rate/AGN luminosity).  
The slope ($\beta$) of $M_{\bullet}-\sigma$ increases for higher thresholds of $SFR$ and $M_{\star}$, indicating that selection effects at these high redshifts (which typically bias samples toward high $SFR$ and $M_{\star}$ objects) are likely to play an important role and lead to biased interpretations of evolutionary effects in these relations when compared to those seen at low-$z$. 
The scatter also tends to increase particularly when high $SFR$ or high $M_{\star}$ samples are selected (which populate the $\sigma \ge 200$ km/s range). 

\begin{figure}
        \begin{subfigure}[t]{\columnwidth}
                \includegraphics[width=\columnwidth]{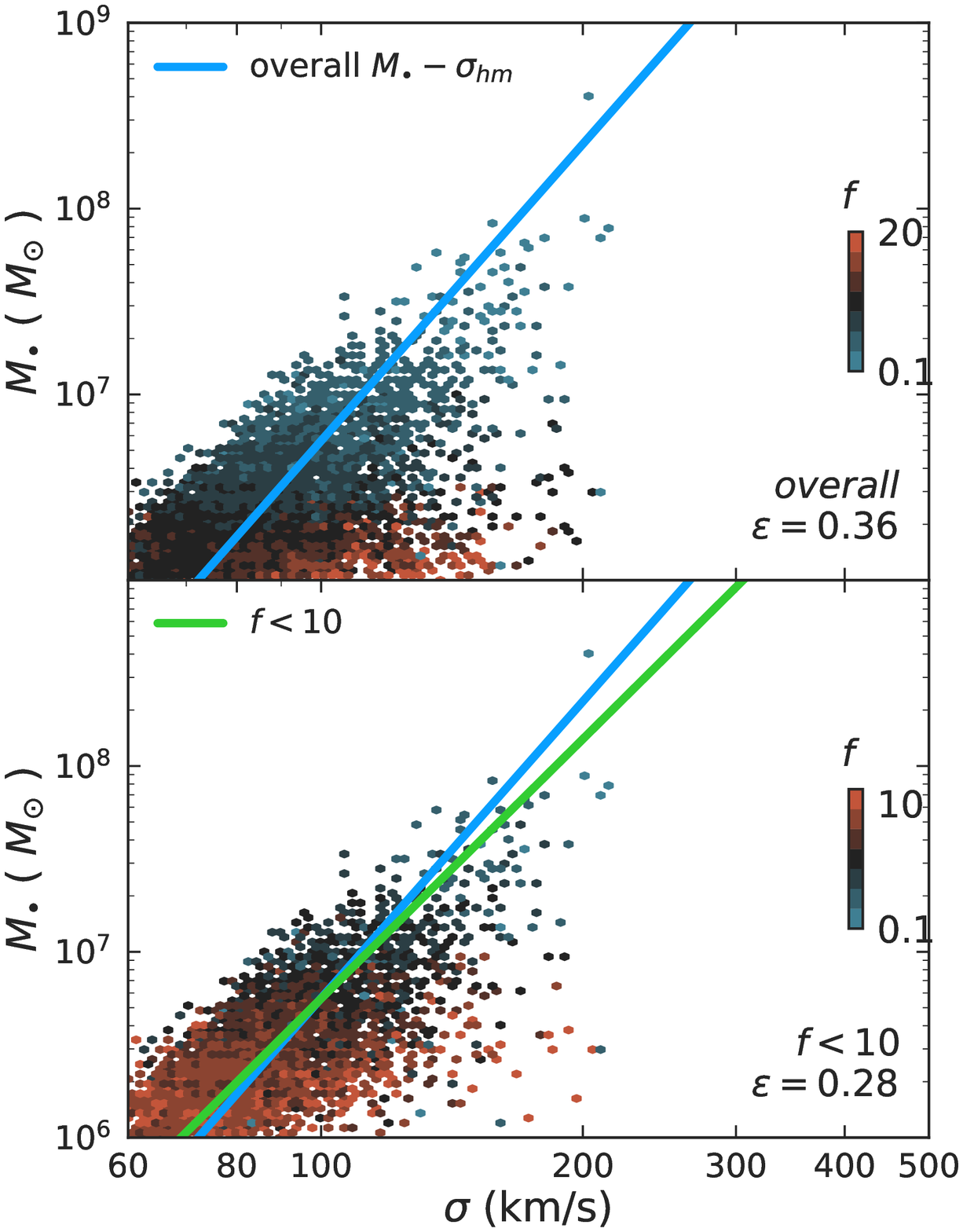}
        \end{subfigure} 
        \begin{subfigure}[t]{\columnwidth}
                \includegraphics[width=\columnwidth]{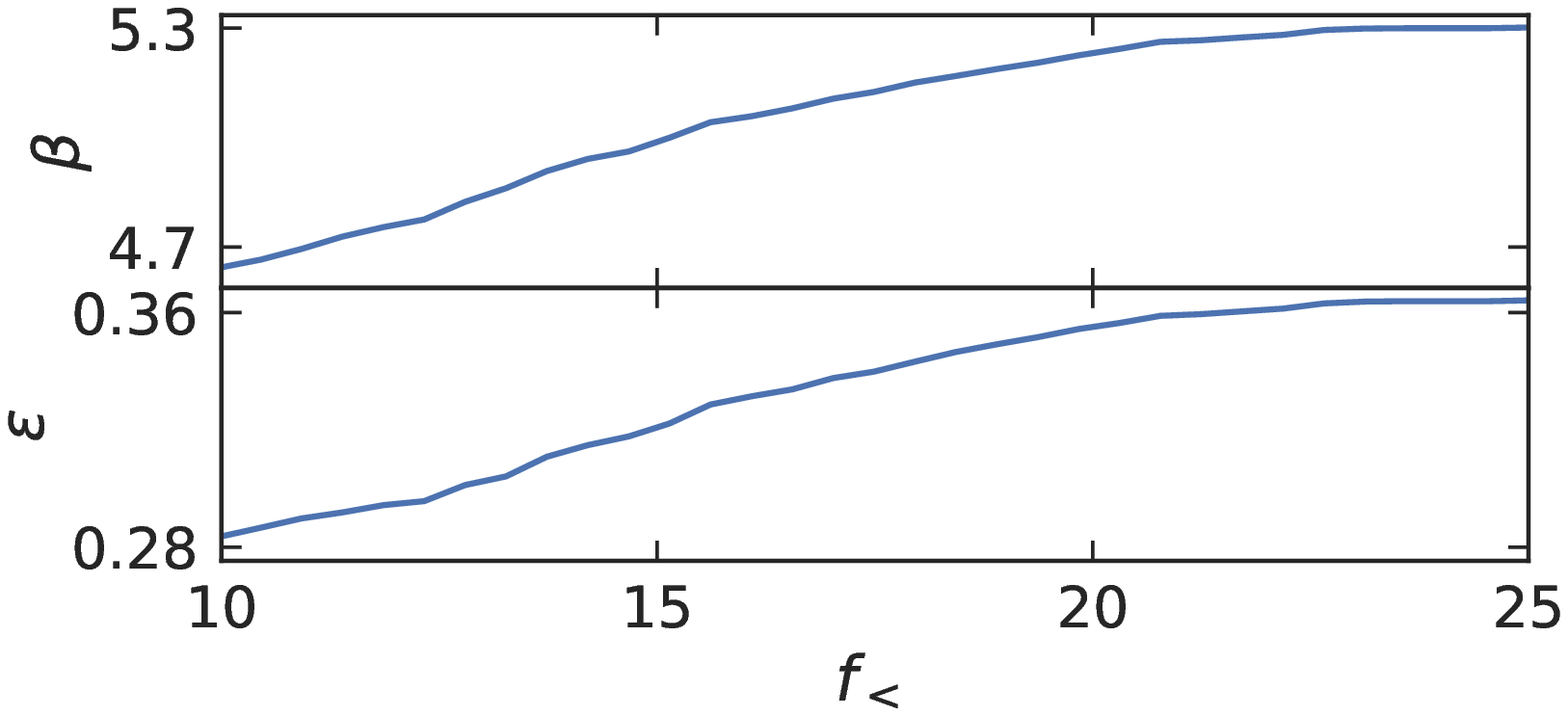}
        \end{subfigure} 
    \caption{Top and middle panels: the $M_{\bullet}-\sigma$ relation at $z=8$ color coded according to the gas-to-stellar ratio ($f$). The blue and green lines are the best fits with overall galaxies and galaxies with $f<10$ respectively. Bottom panel: $\beta$ and $\epsilon$ of $M_{\bullet}-\sigma$ relation at $z=8$ as functions of limiting $f$.} 
    \label{fig:MSgasstar}
\end{figure}

\begin{figure}
    \includegraphics[width=\columnwidth]{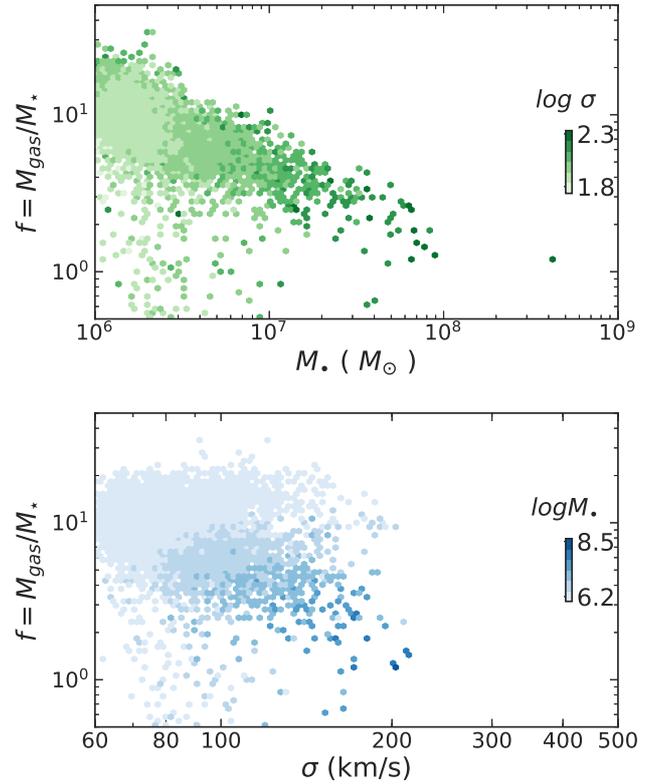}
    \caption{The relations between $f=M_{\rm gas}/M_{\star}$, $M_{\bullet}$, and $\sigma$ color coded according to $\sigma$ and $M_{\bullet}$ (top and bottom respectively).}
    \label{fig:gasstar}
\end{figure}

\subsection{The gas fraction: $f=M_{gas}/M_{\star}$}
\label{sec:gasstellarratio}
In Section~\ref{sec:dependence}, we have examined the dependence of
the scaling relation on the various galaxy or AGN parameters; none of
them helps to explain the relatively large intrinsic scatter in the
$M_{\bullet}-\sigma$ relation. Here, we test for the effects due to
the large range and high value of gas fraction in the high-$z$
galaxies and how that may affect the $M_{\bullet}-\sigma$ relation.

We use the gas-to-stellar ratio ($f$) to measure the gas fraction in
our galaxies in \textsc{BlueTides}. Figure~\ref{fig:MSgasstar} shows
the $M_{\bullet}-\sigma$ relation at $z=8$ color coded according to
$f$.  We find that the gas fraction in galaxies $f$ has a significant
impact on the scatter of the $M_{\bullet}-\sigma$ relation.  In
particular, the scatter is significantly reduced for galaxies with a
smaller value of $f$.  We find that $\epsilon$ decreases significantly
from $0.36$ (all galaxies) to $0.28$ (with $f<10$) while the slope of
the relation ($\beta$) decreases from $5.31$ (all galaxies) to $4.64$
(with $f<10$).  We further illustrate this trend of decreasing of
$\epsilon$ and $\beta$ with different limiting $f$, in the bottom
panel in Figure~\ref{fig:MSgasstar}.  Lower gas fractions are indeed
more representative of local galaxies, which have been used to measure
the $M_{\bullet}-\sigma$.

The decrease of both $\epsilon$ and $\beta$ with the lower limiting
$f$ implies that objects with higher $\sigma$ but lower $M_{\bullet}$
are those that have higher $f$.  To look into the relations between
$f$, $M_{\bullet}$, and $\sigma$, we show Figure~\ref{fig:gasstar},
color coded according to $\sigma$ and $M_{\bullet}$ respectively.  The
top panel indicates a trend between a decreasing $f$ at increasing
$M_{\bullet}$.  Gas fractions of galaxies decrease as BH masses are
high/reach the relation, indicating that BH growth is
quenched/self-regulated due to AGN feedback.  The galaxies with higher
$\sigma$ but lower $M_{\bullet}$ are indeed those that have larger $f$
(the light blue area in the bottom panel), which results in that the
$M_{\bullet}-\sigma$ relation is more scattered if there is no
limiting $f$ applied.  These are relatively sizeable galaxies where
significant BH growth is still occurring (see also
Figure~\ref{fig:GrowthHistory} and objects are still moving toward the
relation (feedback has not yet saturated the BH growth).  Such an
actively growing BH population is indeed rare among the sample of
quiescent local BHs.

\section{The assembly history}
\label{sec:GrowthHistory}

\begin{figure*}
    \includegraphics[width=1.8\columnwidth]{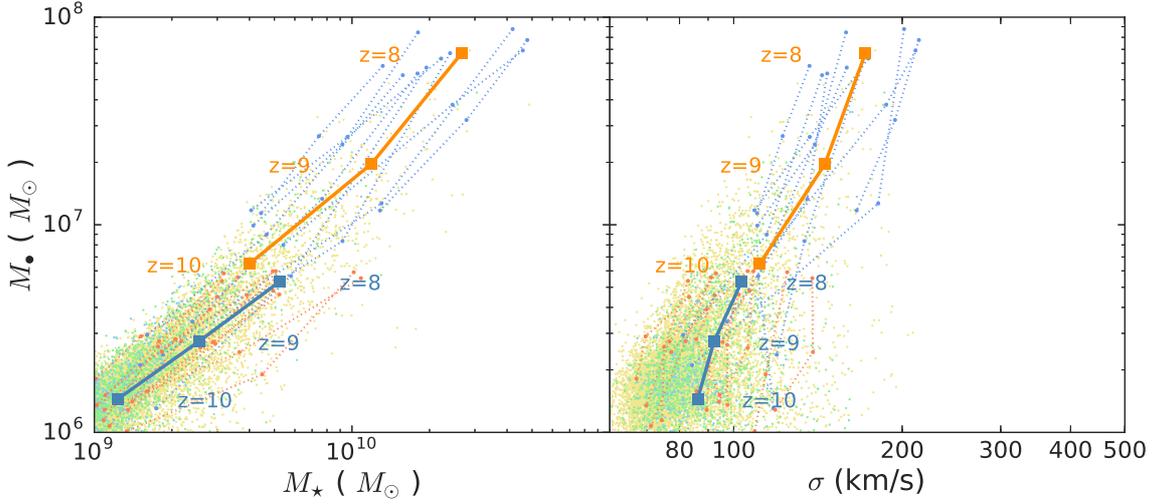}
    \caption{The left and the right panels show the growth history on $M_{\bullet}-M_{\star}$ and $M_{\bullet}-\sigma$ planes of our galaxies from $z=8$ to $z=10$ respectively. The blue and orange dashed curves show the evolution of $\sim10$ galaxies with $M_{\bullet}>5\times10^7M_{\sun}$ and another $\sim10$ galaxies with $M_{\bullet}\sim5\times10^6M_{\sun}$. The blue and orange thick curves demonstrate the average growth history for either groups.}
    \label{fig:GrowthHistory}
\end{figure*}

To further investigate how the  $M_{\bullet}-M_{\star}$ and $M_{\bullet}-\sigma$
relations are established 
we trace the evolution of $\sim 200$ black holes (from $z\sim10$ to
$z\sim8$) and their hosts.  In Figure~\ref{fig:GrowthHistory}, we show
sample tracks on the $M_{\bullet}-M_{\star}$ and $M_{\bullet}-\sigma$
plane from $z=10$ to $z=8$.  The solid lines shown in
orange and blue show the average track in the evolution for higher
mass (with $M_{\bullet}\gtrsim5\times10^7M_{\sun}$) and lower mass
(with $M_{\bullet}\lesssim5\times10^6M_{\sun}$) respectively. The
tracks in the $M_{\bullet}-\sigma$ plane suggest a slightly steeper
growth in the low mass galaxies compared to the high mass galaxies.
This likely indicates a fast growth of the black hole mass at
approximately fixed values of $\sigma$ up to the point where the
galaxies reach the average relation.

To characterize the overall evolution in these planes,
we parameterize the growth of $M_{\bullet}$, $M_{\star}$, and $\sigma$ as
\begin{eqnarray}
\label{eq:growthhistorymm}
    M_{\bullet} \left( z \right) \propto \left( 1+z \right)^{\gamma_{\bullet}} \nonumber \\
    M_{\star}   \left( z \right) \propto \left( 1+z \right)^{\gamma_{\star}} \\
    \sigma      \left( z \right) \propto \left( 1+z \right)^{\gamma_{\sigma}}. \nonumber
\end{eqnarray}
where the exponents are $\gamma_{\bullet} = -9.1 \pm 1.0$,
$\gamma_{\star} = -8.1 \pm 2.2$, and $\gamma_{\sigma} = -1.6 \pm 0.8$,
(the error bars are standard deviation errors).  Note that
$\gamma_{\bullet} / \gamma_{\star} \sim 1.1$ and
$\gamma_{\bullet} / \gamma_{\sigma} \sim 5.7$, which is consistent
with the slope of the scaling relation shown in
Table~\ref{tab:FitCoeffMain} (if we use
Eqs. (\ref{eq:growthhistorymm}) and eliminate $(1+z)$, then
$\gamma_{\bullet}/\gamma_{\star} \sim \beta_{\rm M_{\bullet} -
  M_{\star}}$
and
$\gamma_{\bullet}/\gamma_{\sigma} \sim \beta_{\rm M_{\bullet} -
  \sigma}$).
This suggests that, on average, the redshift evolution of these black
holes traces the overall scaling relation (at a given redshift), with
commensurate growth in black hole mass and stellar mass.

We now take a step further and look into two distinct mass regimes of
the above sample of 200 BHs: 1) A low mass range of BHs with
$M_{\bullet} < 5 \times 10^6 M_{\sun}$, and 2) A high mass range of
BHs with $M_{\bullet} > 1 \times 10^7 M_{\sun}$. We note that, as
discussed in the previous section, the lower mass subsample has more
scatter in the $M-\sigma$ relation (see
Section~\ref{sec:ScalingRelation}).  We find that
$\gamma_{\bullet} =-5.7 \pm 3.5$, $\gamma_{\sigma} = -0.72 \pm 0.49$
and $\gamma_{\bullet} =-10 \pm 2.6$, $\gamma_{\sigma} = -1.8 \pm 0.5$
for low mass and high mass subsamples respectively; therefore, the
higher mass BHs have steeper increase in both $M_{\bullet}$ and
$\sigma$, as is clearly illustrated in Figure~\ref{fig:GrowthHistory}.
Interestingly, we also find
$\gamma_{\bullet} / \gamma_{\sigma} \sim 7.9$, which is higher than
the slope of the overall fit of the $M-\sigma$ relation in
Section~\ref{sec:ScalingRelation}, suggesting that indeed these lower
mass BHs tend to grow their black holes faster than their velocity
dispersion and saturate their growth as they move closer to the mean
relation. This behavior can then potentially lead to a decrease in the
scatter in the relation at the low redshifts, where most black holes
and galaxies have low gas fractions and have quenched their growth and
star formation (particularly in the bulge dominated samples where the
relations are typically measured).  For the high mass sample,
$\gamma_{\bullet} / \gamma_{\sigma} \sim 5.55$ which is consistent
with the slope of the overall fit of the $M-\sigma$ relation, so that
high mass objects continue to move along the relation.

\section{Conclusions}
\label{sec:Conclusions}
We investigate the global properties of supermassive black holes at high redshifts ($z\sim8$, $9$, $10$), which include scaling relations w.r.t properties of their host galaxies ($\sigma$, $M_{\star}$) and their redshift evolution using the \textsc{BlueTides} simulation. 
The bolometric luminosity of BHs span more than two orders of magnitude around a mean of $0.3$ of the Eddington luminosity.
The BH mass functions in our simulation tend to have steeper slopes compared to the one inferred at $z=6$ measured from optical quasars. 
While this may be due to obscuration, we find that it is consistent with the large range of luminosities spanned and the flux cuts implied by observations.
We have also shown that the BH mass density and BH accretion rate are broadly consistent with current observational constraints at the highest redshifts ($z\sim7$).

The scaling relations, $M_{\bullet}-M_{\star}$, $M_{\bullet}-\sigma$ predicted by \textsc{BlueTides} reveal that correlations between the growth of black holes and their host galaxies persist at high-$z$ ($z=8$ to $z=10$), with the slopes and normalizations consistent with published relations at low-$z$. 
For the scatter, we find that the $M_{\bullet}-\sigma$ relation has a significantly higher scatter compared to current measurements as well as the $M_{\bullet}-M_{\star}$ relation. 
We further show that this large scatter can be primarily attributed to the gas-to-stellar ratio ($f=M_{\rm gas}/M_{\star}$), wherein we observe a significant decrease in the scatter ($\epsilon = 0.36$ to $\epsilon = 0.28$) upon exclusion of galaxies with $f=M_{\rm gas}/M_{\star} > 10$. 
Such high gas fraction systems have the largest star formation rates and black hole accretion rates indicating that these fast-growing systems have not yet converged to the relation.

We also find that the assembly history of the evolution of BHs on $M_{\bullet}-M_{\star}$ and $M_{\bullet}-\sigma$ planes is, on an average, consistent with the corresponding scaling relations; in other words, the average trajectory of the evolution of BHs traces the mean scaling relations well.

\section*{Acknowledgements}
We thank A. Tenneti, C. DeGraf, and M. Volonteri for discussions on $D/T$ ratio, the fitting method for scaling relation, and BH mass density respectively. We acknowledge funding from NSF ACI-1614853, NSF AST- 1517593, NSF AST-1616168 and the BlueWaters PAID program. 
The \textsc{BlueTides} simulation was run on the Blue Waters facility at the National Center for Supercomputing Applications.



\bibliographystyle{mnras}
\bibliography{kwbib.bib}



%
%


\bsp	
\label{lastpage}
\end{document}